\documentclass[lettersize,journal]{IEEEtran}
\usepackage{amsmath,amsfonts}
\usepackage{algorithmic}
\usepackage{array}
\usepackage[caption=false,font=normalsize,labelfont=sf,textfont=sf]{subfig}
\usepackage{textcomp}
\usepackage{stfloats}
\usepackage{url}
\usepackage{verbatim}
\usepackage{graphicx}
\def\BibTeX{{\rm B\kern-.05em{\sc i\kern-.025em b}\kern-.08em
    T\kern-.1667em\lower.7ex\hbox{E}\kern-.125emX}}
\usepackage{balance}
\usepackage[nolist]{acronym}
\usepackage{multirow}       
\usepackage{amssymb}        
\usepackage{cite}           
\usepackage[dvipsnames]{xcolor}
\usepackage{soul}
\usepackage[abbreviations]{foreign}     

\usepackage[draft]{pdfcomment} 
\usepackage[hyperfirst=false,acronym,sort=none,shortcuts,nopostdot,style=super,nonumberlist,toc,nogroupskip]{glossaries}

\hypersetup{hidelinks}

\glsdisablehyper 

\newcommand{\accolor}[1]{\textcolor{black}{#1}}

\newacronymstyle{myacro}
{%
  \GlsUseAcrEntryDispStyle{long-short}%
}%
{%
  \GlsUseAcrStyleDefs{long-short}%
}

\setacronymstyle{myacro}

\newcommand*{\tip}[1]{
    \ifglsused{#1}{
      {\pdftooltip{\accolor{\glsentryshort{#1}}}{\glsentrydesc{#1}}}%
    }{%
      \gls{#1}
    }%
}%

\newcommand*{\tips}[1]{
    \ifglsused{#1}{
      {\pdftooltip{\accolor{\glsentryshortpl{#1}}}{\glsentrydesc{#1}}}%
    }{%
      \gls{#1}
    }%
}%

\newacronym{IoT}{IoT}{Internet of Things}
\newacronym{AI}{AI}{Artificial Intelligence}
\newacronym{VAD}{VAD}{Voice Activity Detection}
\newacronym{KWS}{KWS}{Keyword Spotting}
\newacronym{GSCD}{GSCD}{Google Speech Command Dataset}
\newacronym{MEMS}{MEMS}{Micro-Electromechanical Systems}
\newacronym{Mic}{Mic}{Microphone}
\newacronym{ADC}{ADC}{Analog-to-Digital Converter}
\newacronym{MFCC}{MFCC}{Mel Frequency Cepstral Coefficients}
\newacronym{FFT}{FFT}{Fast Fourier Transformation}
\newacronym{SNR}{SNR}{Signal-to-Noise Ratio}
\newacronym{PWM}{PWM}{Pulse-Width Modulation}
\newacronym{PFM}{PFM}{Pulse-Frequency Modulation}
\newacronym{PLL}{PLL}{Phase-Locked Loop}
\newacronym{ADPLL}{ADPLL}{All-Digital Phase-Locked Loop}
\newacronym{AFE}{AFE}{Analog Front-End}
\newacronym{DBE}{DBE}{Digital Back-End}
\newacronym{FEx}{FEx}{Feature Extractor}
\newacronym{FWR}{FWR}{Full-Wave Rectifier}
\newacronym{HWR}{HWR}{Half-Wave Rectifier}
\newacronym{ReLU}{ReLU}{Rectified Linear Unit}
\newacronym{FV}{FV}{Feature Vector}
\newacronym{FVs}{FVs}{Feature Vectors}
\newacronym{VTC}{VTC}{Voltage-to-Time Converter}
\newacronym{SE}{SE}{Single-Ended}
\newacronym{FLL}{FLL}{Frequency-Locked Loop}
\newacronym{DAC}{DAC}{Digital-to-Analog Converter}
\newacronym{VCO}{VCO}{Voltage-Controlled Oscillator}
\newacronym{CCO}{CCO}{Current-Controlled Oscillator}
\newacronym{OSC}{OSC}{Oscillator}
\newacronym{SRO}{SRO}{Switched-Ring-Oscillator}
\newacronym{SRAM}{SRAM}{Static Random Access Memory}
\newacronym{BPF}{BPF}{Band-Pass Filter}
\newacronym{LPF}{LPF}{Low-Pass Filter}
\newacronym{OTA}{OTA}{Operational Transconductance Amplifier}
\newacronym{PD}{PD}{Phase Detector}
\newacronym{PFD}{PFD}{Phase Frequency Detector}
\newacronym{IAF}{IAF}{Integrate-and-Fire}
\newacronym{LUT}{LUT}{Look-Up Table}
\newacronym{DNN}{DNN}{Deep Neural Network}
\newacronym{RNN}{RNN}{Recurrent Neural Network}
\newacronym{CNN}{CNN}{Convolutional Neural Network}
\newacronym{WMEM}{WMEM}{Weight Memory}
\newacronym{HPE}{HPE}{Heterogeneous Processing Element}
\newacronym{FSM}{FSM}{Finite-State Machine}
\newacronym{GRU}{GRU}{Gated Recurrent Unit}
\newacronym{FC}{FC}{Fully Connected}
\newacronym{FoM}{FoM}{Figure of Merit}
\newacronym{D-FF}{D-FF}{D Flip-Flop}
\newacronym{DR}{DR}{Dynamic Range}
\newacronym{TDC}{TDC}{Time-to-Digital Converter}
\newacronym{LSB}{LSB}{Least Significant Bit}
\newacronym{LSTM}{LSTM}{Long Short-Term Memory}
\newacronym{CIC}{CIC}{Cascaded Integrator–Comb}
\newacronym{AGC}{AGC}{Automatic Gain Control}

\begin{document}
\title{A 23 $\mu$W Keyword Spotting IC with Ring-Oscillator-Based Time-Domain Feature Extraction}
\author{Kwantae Kim
, \IEEEmembership{Member, IEEE},
Chang Gao
, \IEEEmembership{Member, IEEE},
Rui Graça
,
Ilya Kiselev
, \IEEEmembership{Member, IEEE},

Hoi-Jun Yoo
, \IEEEmembership{Fellow, IEEE},
Tobi Delbruck
, \IEEEmembership{Fellow}, IEEE,
and Shih-Chii Liu
, \IEEEmembership{Fellow, IEEE}
\thanks{This work was supported in part by the Swiss National Science Foundation, HEAR-EAR, under Grant 200021-172553, and the Ministry of Science and ICT (MSIT), South Korea, through the Information Technology Research Center (ITRC) Support Program supervised by the Institute for Information and Communications Technology Planning and Evaluation (IITP) under Grant IITP-2020-0-01847.
\emph{(Corresponding author: Shih-Chii Liu.)}}
\thanks{Kwantae Kim was with the School of Electrical Engineering, Korea Advanced Institute of Science and Technology (KAIST), Daejeon 34141, South Korea. He is now with the Institute of Neuroinformatics, University of Zürich and ETH Zürich, 8057 Zürich, Switzerland (e-mail: kwantae@ini.uzh.ch).}
\thanks{Chang Gao, Rui Graça, Ilya Kiselev, Tobi Delbruck, and Shih-Chii Liu are with the Institute of Neuroinformatics, University of Zürich and ETH Zürich, 8057 Zürich, Switzerland.}
\thanks{Hoi-Jun Yoo is with the School of Electrical Engineering, Korea Advanced Institute of Science and Technology (KAIST), Daejeon 34141, South Korea.}
\thanks{This paper is accepted for publication in IEEE Journal of Solid-State Circuits (JSSC). \copyright 2022 IEEE. Personal use of this material is permitted. Permission from IEEE must be obtained for all other uses, in any current or future media, including reprinting/republishing this material for advertising or promotional purposes, creating new collective works, for resale or redistribution to servers or lists, or reuse of any copyrighted component of this work in other works.}}
\IEEEaftertitletext{\vspace{-2.0\baselineskip}}

\maketitle

\begin{abstract}
This article presents the first keyword spotting (KWS) IC which uses a ring-oscillator-based time-domain processing technique for its analog feature extractor (FEx). Its extensive usage of time-encoding schemes allows the analog audio signal to be processed in a fully time-domain manner except for the voltage-to-time conversion stage of the analog front-end. Benefiting from fundamental building blocks based on digital logic gates, it offers a better technology scalability compared to conventional voltage-domain designs. Fabricated in a 65\,nm CMOS process, the prototyped KWS IC occupies 2.03mm\textsuperscript{2} and dissipates 23 $\mu$W power consumption including analog FEx and digital neural network classifier. The 16-channel time-domain FEx achieves 54.89\,dB dynamic range for 16 ms frame shift size while consuming 9.3 $\mu$W. The measurement result verifies that the proposed IC performs a 12-class KWS task on the Google Speech Command Dataset (GSCD) with $>$86\% accuracy and 12.4 ms latency.
\end{abstract}


\section{Introduction}\label{sec_intro}
\IEEEPARstart{W}{ith} incredible advances in \tip{AI} fields, there is an increasing demand for low-power audio \tip{IoT} devices that process human speech on the device without data transmission to the cloud. These smart devices are required to ensure always-on operation, real-time response, small form factor, and longer battery lifetime. As such, an ultra-low-power wake-up functionality is being highlighted with rapidly growing popularity because it allows hierarchical power gating of increasingly complex tasks for audio \tip{IoT} nodes. \tip{KWS} and \tip{VAD} are widely used user-interactive methods to wake-up smart devices. \tip{KWS} is used to detect predefined keywords in an audio stream while \tip{VAD} detects when a human voice is present.

\begin{figure}[t]
    \begin{center}
    \includegraphics[width=\columnwidth]{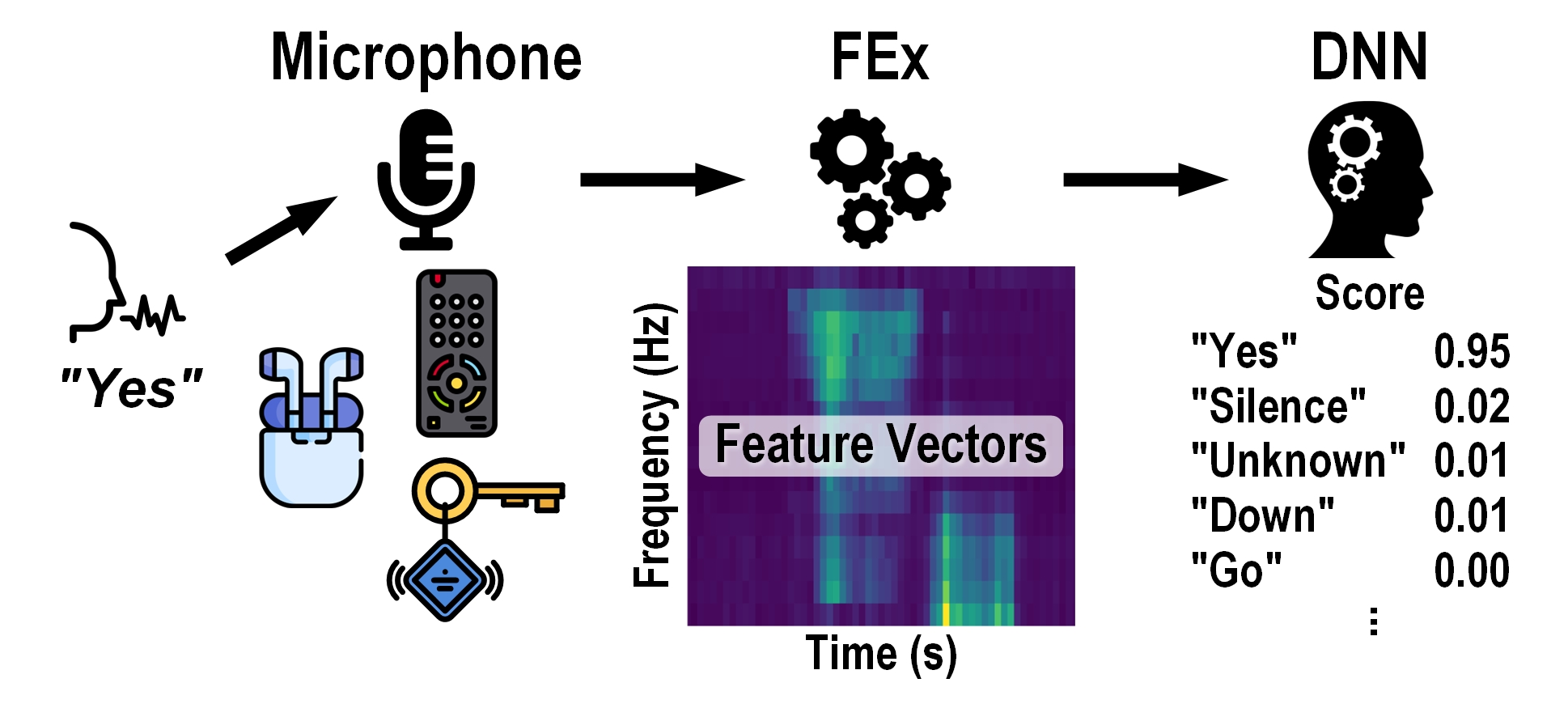}
    \caption{Processing stages for \tip{KWS} in an audio \tip{IoT} device.}\label{fig:intro}
    \end{center}
    \vspace{-5mm}
\end{figure}

Fig. \ref{fig:intro} shows the typical processing stages for \tip{KWS}. The user says a keyword into the microphone of an edge device such as a remote control or wireless earbud. The microphone output is further processed by a \tip{FEx} which generates frequency-selective \tip{FVs} that are continuously streamed to a \tip{DNN}-based classifier. The classifier outputs the probability scores of different keywords. \tip{IoT} devices benefit from a tiny form factor and the use of a small battery such as a coin cell, e.g., for smart tags. Generally a $<$100\,$\mu$W system-level power is desirable including not only the \tip{KWS} IC itself but also the microphone and other system components. Moreover, a low-latency response is desired considering a \tip{KWS}-driven hierarchical processing system used in an interactive environment. For example, a study on the perception of self-generated speech showed that a delay exceeding 20\,ms becomes disturbing for users \cite{stone1999tolerable}.

\begin{figure*}[t]
    \begin{center}
    \includegraphics[width=\textwidth]{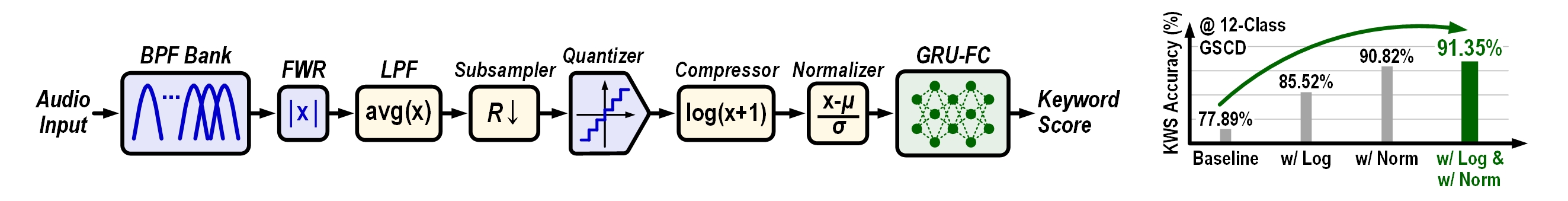}
    \caption{Architecture of the KWS software model (left) and simulated KWS accuracy (right).}\label{fig:software}
    \end{center}
    \vspace{-5mm}
\end{figure*}

A 12-class \tip{KWS} IC that includes the whole processing chain starting from the \tip{ADC} to the \tip{DNN} classifier \cite{giraldo2019kws} reported that the \tip{FEx} is the most power-hungry stage accounting for 40\% of the power dissipation in the entire IC. To reduce the power budget of edge devices thereby facilitating longer battery lifetime or smaller battery size, various circuit design techniques have been proposed for both \tip{KWS} and \tip{VAD} ICs. However, most of them traded off between power and latency. In \cite{oh2019vad}, a 142\,nW \tip{VAD} IC using sequential mixer-based \tip{FEx} was proposed where the operational principle is similar to that used for bio-impedance sensors \cite{kim2020bioz, ha2019bioz}. However, this sequential frequency scanning is too undersampled for the \tip{KWS} and results in a 512\,ms latency for \tip{VAD}. In \cite{shan2021kws}, a serialized digital \tip{FEx} was used in the \tip{KWS} IC where the processing stages are pipelined. Although this IC consumed only 510\,nW, its latency was limited to 64\,ms and it needed an off-chip 16-bit \tip{ADC}, had only 2KB memory for binary \tip{CNN} thereby its accuracy was only reported for 5 keywords.

Another approach is the use of an analog voltage-domain \tip{FEx} which exploits low-power analog circuits to achieve both low-power and low-latency responses. The processing chain of the analog \tip{FEx} typically consists of a multi-channel \tip{BPF}, a \tip{HWR} or a \tip{FWR}, and an \tip{ADC}. Here, the speed requirement of \tip{ADC} is highly relaxed to 10\,ms-to-100\,ms (10\,Hz-to-100\,Hz), which corresponds to the size of frame shift in audio signal processing. This is possible because the output of rectifier represents the magnitude response of the input speech and thus it is a low-frequency signal. Previous works that used a voltage-domain analog \tip{FEx} and a back-end classifier to implement \tip{VAD} \cite{badami2016vad, yang2019vad} and \tip{KWS} \cite{wang2021kws} tasks, reported 205\,nW-to-1\,$\mu$W power dissipation and 10\,ms-to-100\,ms latency. However, voltage-domain analog \tip{FEx} is unfriendly for CMOS technology scaling, thereby the power efficiency of analog approaches are predicted to be degraded in advanced nanometer-scale processes. This is because $V_\text{DD}$ is scaling down faster than $V_\text{TH}$, thus voltage-domain signals have less headroom. Reduced headroom results in reduced maximal signal swing, which in turn, reduces the \tip{DR} that is critical for keeping \tip{KWS} accuracy high across a range of audio amplitude levels. Furthermore, the intrinsic gain ($g_\text{m}r_\text{o}$) of the transistors is also degraded, leading to the DC gain reduction in analog feedback loops. This issue can be mitigated with a larger transistor length, gain boosting, or multi-stage amplifiers; however, these approaches come with costs in area, power, and bandwidth.

To this end, we propose a time-domain analog \tip{FEx} that exploits the scaling-friendly nature of the ring-oscillator. It is the first silicon-verified ring-oscillator-based audio \tip{FEx} reported to date. When integrated with an on-chip \tip{RNN} classifier, the resulting IC demonstrates power-efficient \tip{KWS} capability. The \tip{FEx} circuits extensively use time-domain signal representation techniques including \tip{PWM} and \tip{PFM}, therefore it does not suffer from headroom degradation and its associated signal swing loss issue. In other words, it is more suitable for low-supply implementation than voltage-domain designs. The ring-oscillator-based circuit utilizes its infinite DC gain characteristic when configured as a time-domain integrator \cite{drost2012ringfilter}. As such, the transfer function of time-domain \tip{FEx} circuits such as \tip{BPF} are not affected by the degradation of the intrinsic gain of transistors. Overall, the proposed \tip{KWS} IC consumes 23\,$\mu$W and {has only} 12.4\,ms {inference} latency on a 12-class \tip{GSCD}~\cite{gscd}.

There have been similar approaches to implement the oscillator-based \tips{BPF} for audio \tip{IoT} applications \cite{Gutierrez2019vcofilter, Goux2020ilofilter}. However, none of them proposed a clear design strategy to implement a time-domain rectifier or demonstrated an audio classification task using the fabricated oscillator-based \tips{BPF}.

This article is an extension of a previous work presented in \cite{kim2022kws}. The integrated chip also includes a switched-capacitor energy harvester circuit, a voltage reference, and a low-dropout regulator. However, {in this paper,} we focus on the new circuit techniques of the \tip{KWS} core. The paper is organized as follows. Section~\ref{sec_software} presents the software modeling of the \tip{KWS} modules in this work. Section~\ref{sec_KWS_IC} covers the description for the overall architecture and design details of the implemented circuits and Section\,\,\ref{sec_measurement} presents measurement results and performance summary of the prototype chip. Section\,\,\ref{sec_conclusion} concludes this work.

\begin{figure*}[t]
    \begin{center}
    \includegraphics[width=\textwidth]{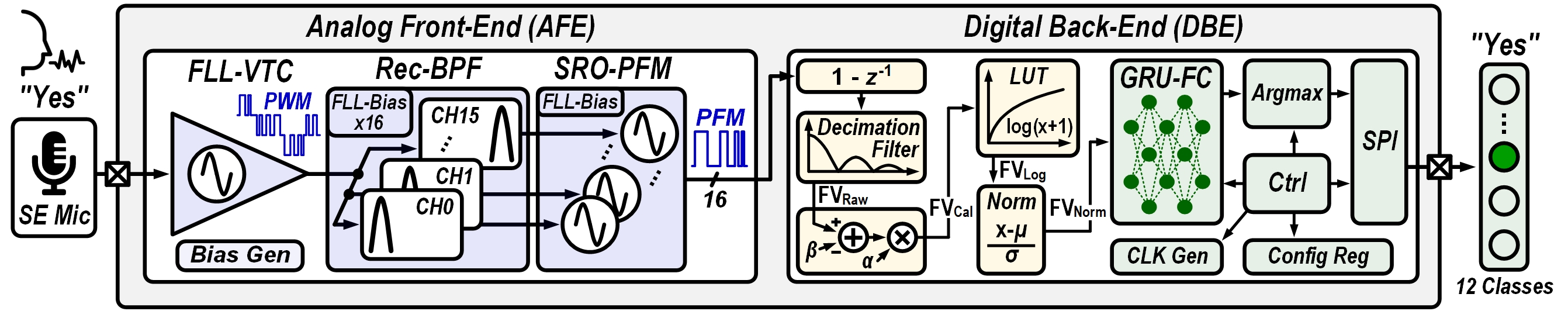}
    \caption{Overall architecture of the proposed KWS IC.}\label{fig:overall}
    \end{center}
    \vspace{-5mm}
\end{figure*}

\begin{figure}[t]
    \begin{center}
    \includegraphics[width=\columnwidth]{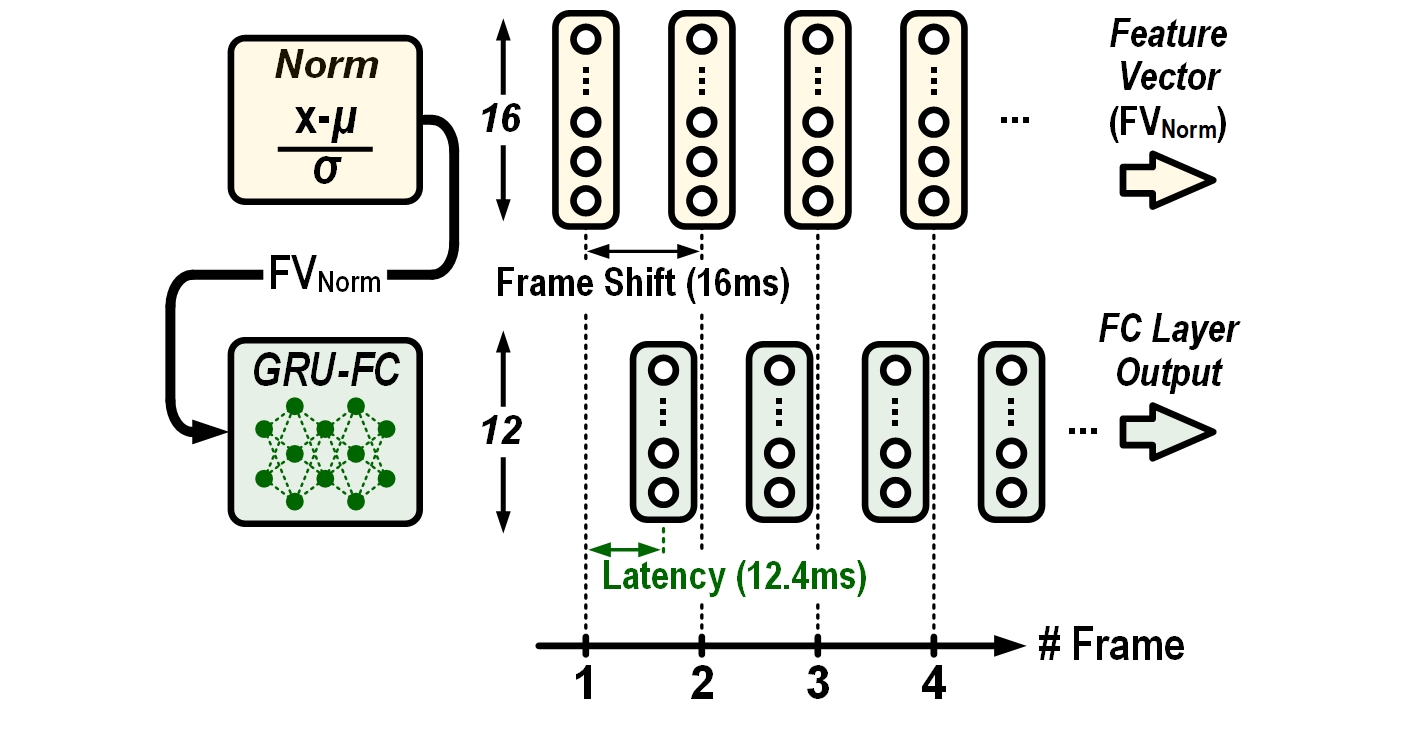}
    \caption{Timing diagram of the GRU-FC classifier computation according to the input feature vectors.}\label{fig:timing}
    \end{center}
    \vspace{-5mm}
\end{figure}

\section{Software Modeling}\label{sec_software}

The architecture of our \tip{KWS} IC was developed based on prior silicon cochlea and edge audio-inference ICs \cite{yang2019vad, badami2016vad, wang2021kws, kim2022review, lyon1988cochlea}. We implemented a Python model of the \tip{KWS} IC including the analog \tip{FEx} as shown in Fig.~\ref{fig:software}. Our model implements a bank of \tips{BPF} (second-order butterworth filter) inspired by modeling of biological cochlea \cite{lyon1988cochlea}, a \tip{FWR} ($|x|$), an averaging block (low-pass filter), a subsampler, and a quantizer. The subsampler was added to realize the relaxed speed requirement of the quantizer as discussed in Section~\ref{sec_intro}. As the \tip{GSCD} samples have a 16\,kHz sampling rate, the number of averaged samples and the rate of subsampling operation were selected to match the target frame shift size (16\,ms in our work) of the audio \tip{FV}. In contrast to prior analog \tips{FEx} \cite{yang2019vad, badami2016vad}, we added additional \tip{FV} processing stages before the \tip{FV} is fed to the classifier. These stages consist of 1) a logarithmic compression stage inspired by the adaptive gain compression mechanism of biological cochleas and 2) an input normalization stage which is widely used in \tip{DNN} models, both of which help to improve the \tip{KWS} accuracy on \tip{GSCD}. We chose a \tip{GRU}-based \tip{RNN} classifier for the last stage of our \tip{KWS}, as it has been frequently used in automatic speech recognition tasks \cite{amodei2016DeepSpeech}.

Fig.~\ref{fig:software} shows the accuracy of the software simulation starting from the baseline model which does not include the compressor and normalizer stages. As seen on the right graph of Fig.~\ref{fig:software}, the baseline model achieved 77.89\% which increases to 91.35\% KWS accuracy on the 12-class \tip{GSCD} test set with the addition of the 2~stages. The following design parameters were chosen for our software model: First, we used a 16-channel \tip{BPF} which was also used in previous works \cite{yang2019vad, badami2016vad, wang2021kws}, and a $Q$-factor of 2 for the \tips{BPF}. Second, the center frequencies of the bank of \tips{BPF} are distributed according to the Mel scale (from 100\,Hz to 8\,kHz). The 8\,kHz value is also the bandwidth of the analog front-end presented in \cite{giraldo2019kws}. Note that we oversampled the input speech 2$\times$ (from 16\,kHz to 32\,kHz sampling rate), to avoid the 8\,kHz center frequency overlapping with Nyquist frequency (8\,kHz with a 16\,kHz sampling rate). Third, a 12-bit quantizer (before logarithmic compression in Fig.~\ref{fig:software}), Fourth, a 16\,ms frame shift, the same value used in \cite{giraldo2019kws, shan2021kws}, Fifth, a 10-bit output logarithmic compressor and a 14-bit normalized feature vector ($\text{FV}_\text{Norm}$) that is fed to a 2-layer 48-hidden-unit \tip{GRU} and an \tip{FC} layer. Sixth, 14-bit and 8-bit quantizations were applied to the activations and weights respectively. The baseline model accuracy shown in Fig.~\ref{fig:software} would be higher if using floating-point activations because the 14-bit quantization (6-bit integral part and 8-bit fractional part) cannot cover the dynamic range of the 12-bit unsigned quantizer output.

\section{KWS IC with Time-Domain Analog FEx}\label{sec_KWS_IC}

The overall architecture of our \tip{KWS} IC is shown in Fig. \ref{fig:overall}. It is composed of an analog front-end and a digital back-end. The analog front-end is designed to match our software model. The first stage of the analog front-end is a \tip{FLL}-based \tip{VTC} (Section~\ref{sec_vtc}). It features a nested analog \tip{FLL} circuit that linearizes the voltage-to-frequency response of the \tip{VCO}. A voltage-domain audio input from a \tip{SE} \tip{Mic} is converted into a time-domain multi-phase \tip{PWM} output through the \tip{VTC}. The second stage of the analog front-end is a 16-channel rectifying \tip{BPF} (Rec-\tip{BPF}). Each channel has a time-domain second-order \tip{BPF} (Section~\ref{sec_bpf}) featuring an inherent \tip{FWR} functionality. The output of each \tip{BPF} channel are \tip{PWM} signals and they are further converted into \tip{PFM} signals through a \tip{SRO}-based rate-encoder.

\begin{figure*}[t]
    \begin{center}
    \includegraphics[width=\textwidth]{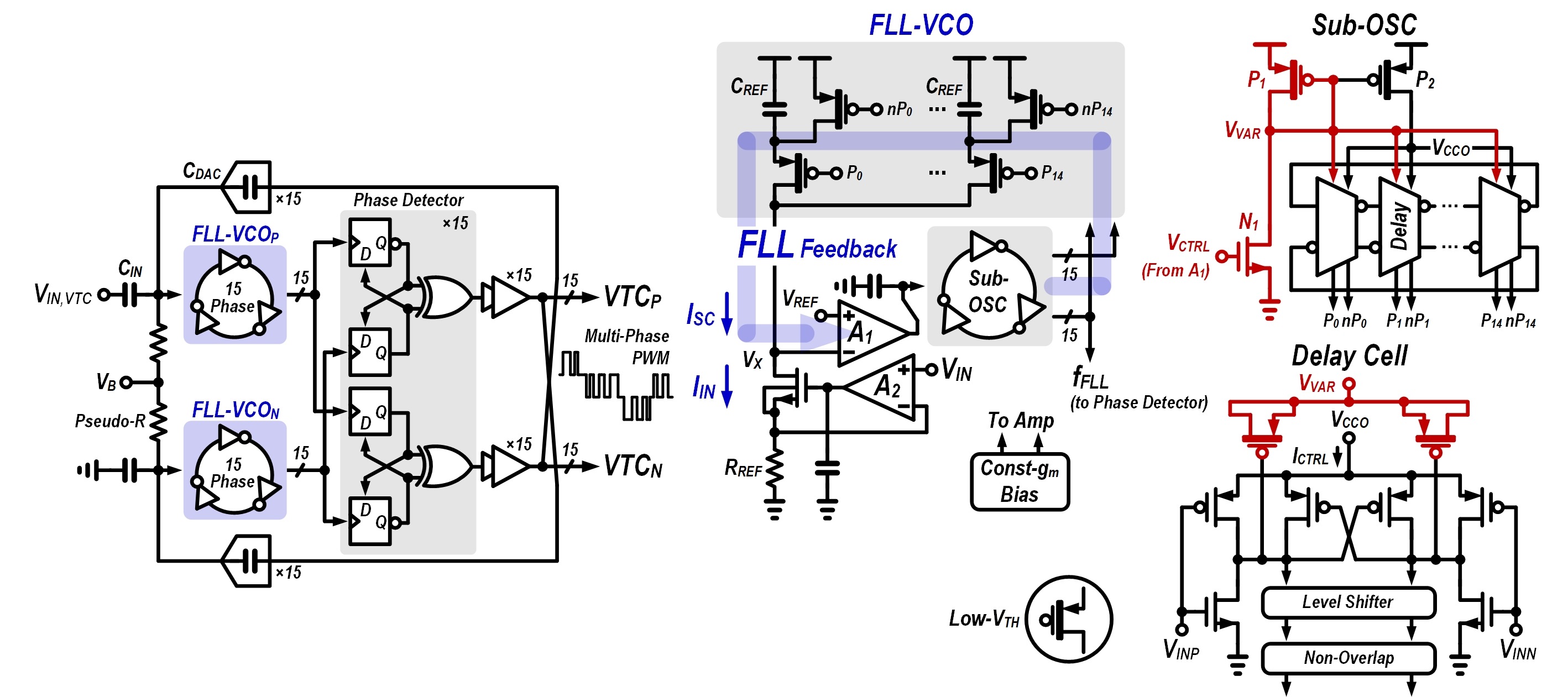}
    \caption{0.5 V supply single-ended (SE) input \tip{FLL}-based \tip{VTC}.}\label{fig:vtc}
    \end{center}
    \vspace{-5mm}
\end{figure*}

\begin{figure}[t]
    \begin{center}
    \includegraphics[width=\columnwidth]{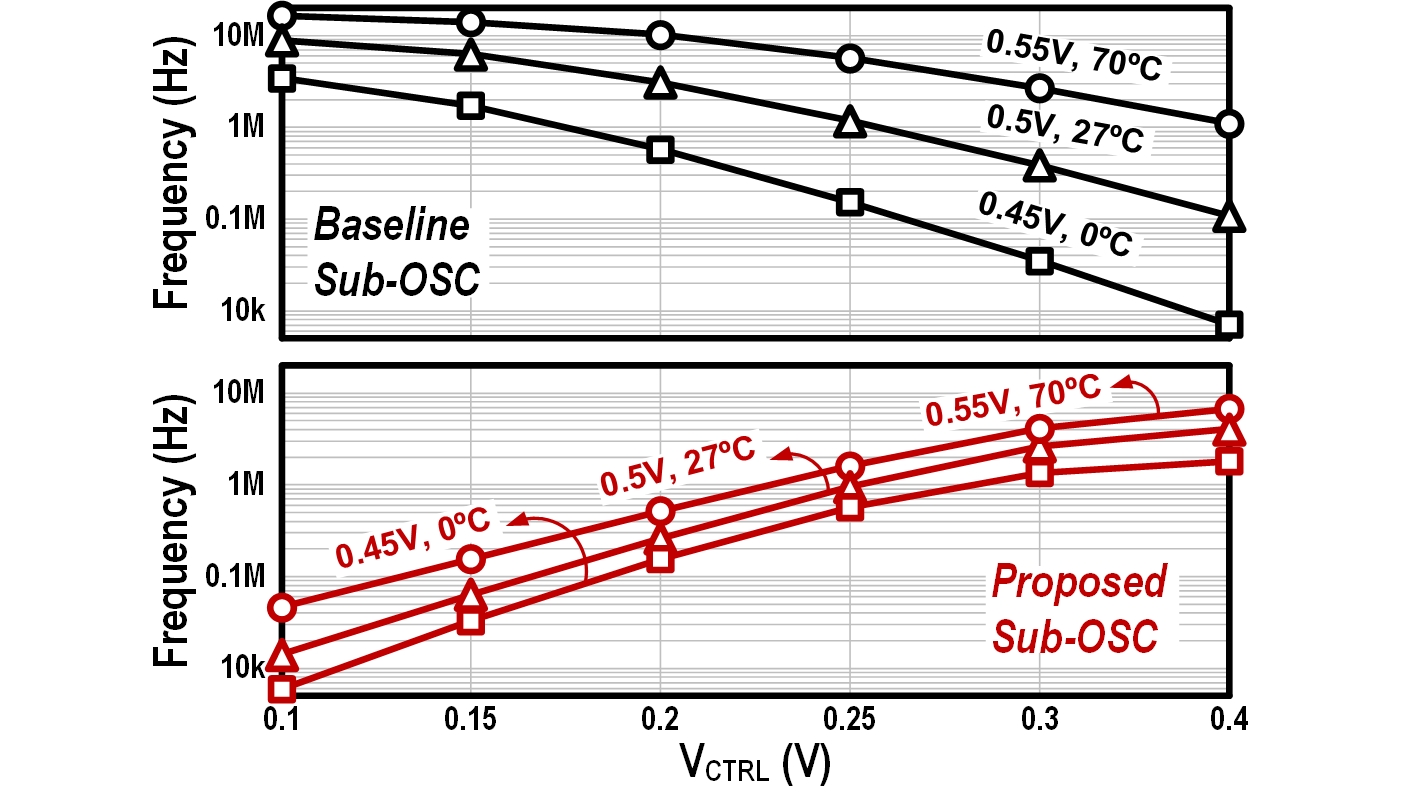}
    \caption{Simulation result of the supply-temperature compensation in sub-\tip{OSC}.}\label{fig:vtc_sub-osc_sim}
    \end{center}
    \vspace{-5mm}
\end{figure}

\begin{figure}[t]
    \begin{center}
    \includegraphics[width=\columnwidth]{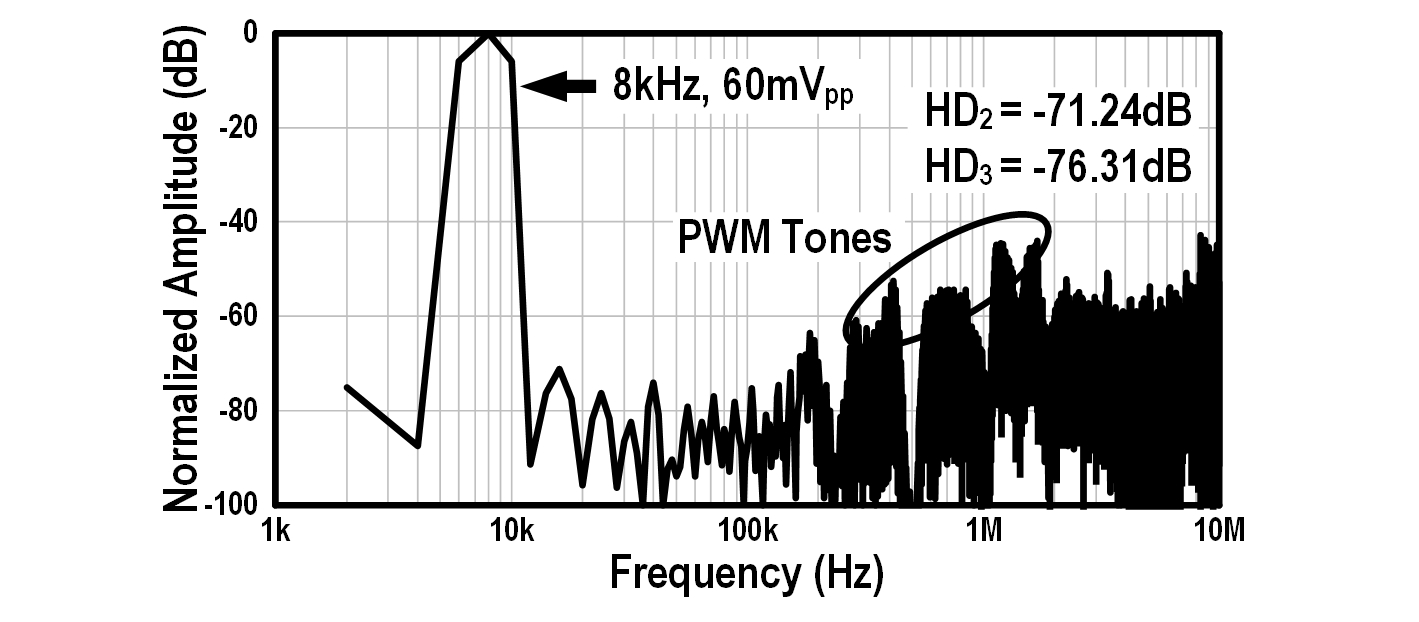}
    \caption{Simulation result of the \tip{FLL}-based \tip{VTC}.}\label{fig:vtc_sim}
    \end{center}
    \vspace{-5mm}
\end{figure}

In the digital back-end, the \tip{PFM} signals are fed into a digital differentiator ($1-z^{-1}$). Here, the signal path from the \tip{SRO} to the digital differentiator builds a first-order $\Delta\Sigma$ \tip{TDC} \cite{elshazly2014sro} which corresponds to the quantizer in our software model. The output of the digital differentiator is further processed through subsequent stages including a decimation filter which performs the averaging and subsampling operation in our software model. It also includes an offset subtractor ($\beta$) which removes the free-running frequency component of the \tip{SRO}, a per-channel gain calibrator ($\alpha$) which corrects the inter-channel gain mismatch, a logarithmic \tip{LUT}, and an input normalizer. Both $\mu$ and $\sigma$ shown in Fig.~\ref{fig:overall} are respectively the mean and the standard deviation of the output of logarithmic \tip{LUT} ($\text{FV}_\text{Log}$ in Fig.~\ref{fig:overall}) from our chip with the \tip{GSCD} training set. With the normalizer, $\mu$ is subtracted from the $\text{FV}_\text{Log}$ and the resulting subtracted output is multiplied by a value $1/\sigma$. The resulting output of the $\text{FV}_\text{Norm}$ is a 16-channel signed 14-bit \tip{FV} which is generated every 16\,ms of frame shift as shown in Fig.~\ref{fig:timing}. For each \tip{FV}, a 2-layer \tip{GRU} \tip{RNN} and 1-layer \tip{FC} digital accelerator outputs the most probable keyword over 12 classes with a 12.4\,ms latency (see Fig.~\ref{fig:timing}).

\subsection{Voltage-to-Time Converter}\label{sec_vtc}

Previous \tip{VAD} and \tip{KWS} ICs have used mainly a differential output microphone interface \cite{giraldo2019kws, yang2019vad, wang2021kws}. However, commercial off-the-shelf differential output \tip{MEMS} microphones typically consume $>$100 $\mu$W. To realize a system-level low-power audio \tip{IoT} device, a low-power \tip{SE}-interface \tip{MEMS} microphone is preferred because it consumes as little as $\sim$10\,$\mu$W (e.g., InvenSense ICS-40310 \cite{ics-40310}). But this approach makes it difficult to obtain good linearity because \tip{SE} signals do not reject even-order harmonics. In general, a linear \tip{FEx} is preferred as it makes training the back-end \tip{DNN} classifier easier and enhances the spectral purity of an audio signal with minimal harmonics and intermodulation distortions. Particularly, the design of a \tip{SE}-input ring-oscillator-based \tip{VTC} circuit becomes even more difficult because \tips{VCO} exhibit poor linearity compared to a conventional voltage-domain \tips{OTA}.

In this work, we propose to use a nested analog \tip{FLL} around the ring-\tip{VCO} to enhance the linearity of the \tip{VTC}. Fig. \ref{fig:vtc} shows the architecture and transistor-level schematic of the \tip{FLL}-based \tip{VTC}. The fundamental design principle is adopted from the ring-oscillator-based \tip{LPF} presented in \cite{drost2012ringfilter}, however capacitive coupling is used with $C_\text{IN}$ to isolate the DC bias of the \tip{VTC} from the microphone. Therefore, its core operation is similar to the capacitively-coupled voltage amplifier \cite{Harrison2003CCIA}, but the \tip{VTC} circuit converts the input voltage into the multi-phase \tip{PWM} output instead of voltage. A pseudo-differential architecture is implemented using a dual-\tip{VCO} structure along with the phase detector \cite{drost2012ringfilter}. One input port of the \tip{VTC} is connected to the \tip{SE} microphone and the other input port is tied to ground. The 15-array phase detector receives a 15-phase frequency-modulated signal out of the \tips{VCO} and generates a 15-phase \tip{PWM} output which represents the phase difference of the \tips{VCO}. Note that exploiting the multi-phase \tip{PWM} scheme pushes spurious \tip{PWM} tones to higher frequency range without necessitating a higher running frequency of the \tip{VCO} \cite{drost2012ringfilter}. The outputs of the phase detector are buffered with two inverters and used to close the feedback loop through a 15-array thermometer-coded capacitive \tip{DAC}. Since the input node of the \tip{VCO} acts as a virtual-ground, the generated multi-phase \tip{PWM} signal becomes a time-domain approximated input voltage where the amplitude is encoded into the duty-cycle of \tip{PWM}. A variation-tolerant pseudo-resistor \cite{djekic2018pseudores} with a voltage reference $V_\text{B}$ sets the common-mode DC bias voltage of the \tips{VCO}.

The \tip{FLL}-based \tip{VCO} includes a single-branch current comparator \cite{jang2018pll} to operate the analog \tip{FLL}. As shown in the schematic diagram of the \tip{FLL}-\tip{VCO} in Fig. \ref{fig:vtc}, an input current generator drives the input voltage to $R_\text{REF}$ to generate a low-side current signal $I_\text{IN}=V_\text{IN}/R_\text{REF}$ where $A_{2}$ amplifier is designed to have a 34\,dB gain. A high-side current $I_\text{SC}=15V_\text{X}C_\text{REF}f_\text{FLL}$ flows through a 15-phase switched-capacitor operation. Here, the multi-phase nature of a ring-oscillator is fully utilized to apply the multi-phase interleaving technique at the $V_\text{X}$ node to minimize the voltage ripple caused by the switched-capacitor operation \cite{jang2018pll}. The low-$V_\text{TH}$ devices are used to facilitate 0.5 V low-supply operation for the implementation of the switched-capacitor circuit. The \tip{FLL} feedback formed through the $A_{1}$ amplifier with a 27\,dB gain, sub-\tip{OSC}, and switched-capacitor circuit ensures that $V_\text{X}$ equals the reference voltage $V_\text{REF}$ while also ensuring $I_\text{IN}$ equals $I_\text{SC}$. As a result, the output frequency of the \tip{FLL}-based \tip{VCO} is set as in \eqref{eq:fll}.
\begin{align}
    f_\text{FLL}&=\frac{V_\text{IN}}{15R_\text{REF}C_\text{REF}V_\text{REF}} \label{eq:fll}\\
    K_\text{FLL-VCO}&
    =\frac{\partial f_\text{FLL}}{\partial V_\text{IN}}
    =\frac{1}{15R_\text{REF}C_\text{REF}V_\text{REF}}
    \label{eq:K_fll}
\end{align}

Since $f_\text{FLL}$ is represented by the input voltage $V_\text{IN}$ and reference parameters such as $R_\text{REF}$, $C_\text{REF}$, and $V_\text{REF}$, it leads to a \tip{FLL}-aided linearization of the \tip{VCO} as derived in (\ref{eq:K_fll}), where $K_\text{FLL-VCO}$ corresponds to the voltage-to-frequency tuning gain. This is because the value of passive elements ($R_\text{REF}$, $C_\text{REF}$, and $V_\text{REF}$) has no dependency on the input signal amplitude ($V_\text{IN}$).

The 3\,dB bandwidth of the \tip{VTC} circuit is given as below, which is similar to the equation of resistive-input and current-feedback ring-oscillator-based filter \cite{drost2012ringfilter}
\begin{align}
    f_\text{3dB,VTC}=\frac{1}{2\pi}K_\text{FLL-VCO}K_\text{PD}\beta_\text{DAC}\\
    \beta_\text{DAC}=\frac{15C_\text{DAC}}{C_\text{IN}+15C_\text{DAC}}V_\text{DD}
\end{align}
where $K_\text{PD}$ is the gain of phase detector and $\beta_\text{DAC}$ is the feedback factor (time-to-voltage). We designed $f_\text{3dB,VTC}$ to be 17\,kHz when the nested \tip{FLL} feedback has a 158\,kHz gain-bandwidth product which contributes as a non-dominant pole to the overall negative feedback loop of the \tip{VTC} circuit. As shown in Fig.~\ref{fig:vtc_sim}, the stability of the \tip{VTC} is verified with a transient simulation.

To allow 0.5\,V low-supply operation for the sub-\tip{OSC}, a varactor-controlled supply-temperature compensator is proposed. This is achieved by sizing the diode-connected transistor $P_{1}$ so that $V_\text{VAR}$ becomes proportional-to-absolute-temperature (PTAT) \cite{seok2012vref}. Therefore, the capacitance of MOS-varactors in the delay cells \cite{Zhao2020VCOADC} adaptively stabilizes the temperature drift of the ring-oscillator frequency. For example, if the temperature increases, then $V_\text{VAR}$ also increases, therefore the MOS-varactors are further turned on. This effect negates the frequency increase of the ring-oscillator with temperature increase. Low-$V_\text{TH}$ MOS capacitors are used to further enhance the varactor effect. In addition, instead of configuring the sub-\tip{OSC} as controlled by the gate voltage of $P_{2}$ only, the $N_{1}-P_{1}$ path is added to reduce $V_\text{DD}$ sensitivity based on the fact that $V_\text{GS,N1}$ is less sensitive to $V_\text{DD}$ than $V_\text{SG,P2}$. The simulation results in Fig. \ref{fig:vtc_sub-osc_sim} show that with the proposed techniques, the supply-temperature variation of the sub-\tip{OSC} is reduced by 19.98$\times$ in the worst case. Note that the baseline sub-\tip{OSC} refers to the \tip{OSC} circuit assuming that the added compensation circuits (marked as red color in Fig. \ref{fig:vtc}) are removed. In this case, $V_\text{CTRL}$ is connected to the gate of the $P_{2}$ transistor and therefore the frequency tuning curve of sub-\tip{OSC} becomes decreasing function as $V_\text{CTRL}$ increases. Fig.~\ref{fig:vtc_sim} shows the post-layout simulation result of the \tip{VTC}. The plotted graph represents the multi-phase \tip{PWM} signal of \tip{VTC} output ($VTC_\text{P}-VTC_\text{N}$). The designed \tip{VTC} converts a voltage-domain input into a time-domain \tip{PWM} output while ensuring $<$-70 dB distortion for dominant harmonics (second and third) even with a \tip{SE} input. The \tip{PWM} tones at higher frequencies are filtered out at the following \tip{BPF} stage.

\begin{figure}[t]
    \begin{center}
    \includegraphics[width=\columnwidth]{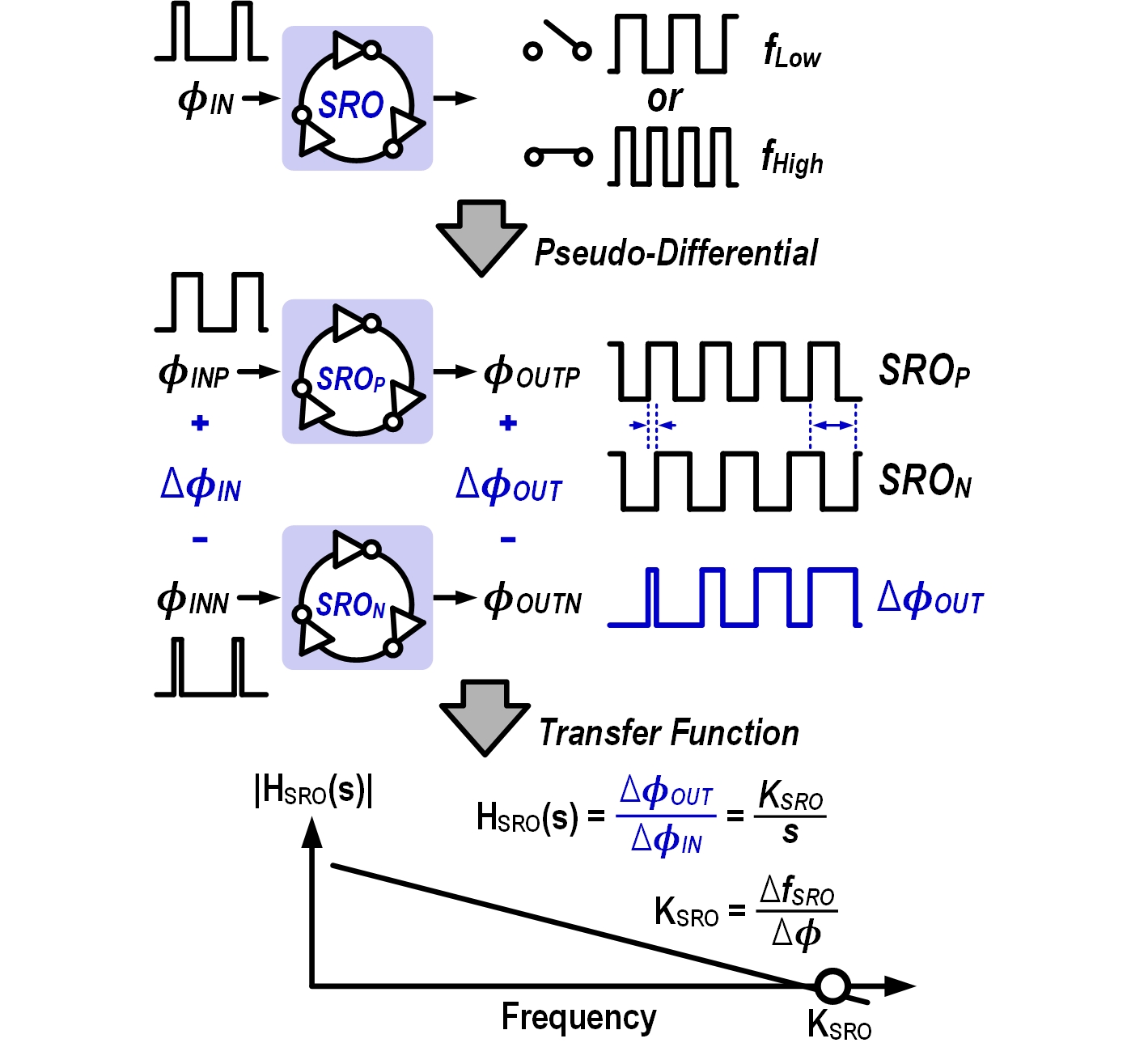}
    \caption{\tip{SRO} as an ideal $\phi$-to-$\phi$ integrator.}\label{fig:sro-integ}
    \end{center}
    \vspace{-5mm}
\end{figure}

\subsection{Time-Domain Band-Pass Filter}\label{sec_bpf}

Fig. \ref{fig:sro-integ} shows a conceptual diagram of using the switched-ring-oscillator (SRO) \cite{elshazly2014sro} as an ideal $\phi$-to-$\phi$ integrator. The \tip{SRO} switches its running frequency between $f_\text{Low}$ and $f_\text{High}$ according to the incoming input \tip{PWM} signal. The averaged value of \tip{SRO} frequency is proportional to the duty-cycle of the input \tip{PWM} signal. If the input \tip{PWM} signal is configured as a multi-phase format, the possible number of running frequencies also increases. When the dual-\tip{SRO} is implemented in a pseudo-differential manner, the output phase difference $\Delta\phi_{OUT}$ becomes an accumulated (or integrated) input phase difference $\Delta\phi_{IN}$ over time. Specifically, this integral procedure flows as following. \textit{Input Phase} $\rightarrow$ \textit{SRO Frequency} $\rightarrow$ \textit{SRO Phase (+Integral)} $\rightarrow$ \textit{Output Phase}. The phase mathematically represents an integral amount of the frequency within an oscillator. This time-domain accumulation process allows an integral of the signal without boundary as long as the \tip{SRO} oscillates, unlike voltage-domain designs that saturate due to headroom. In other words, it shows an infinite DC gain and acts as a true lossless integrator regardless of the intrinsic gain of transistor or supply voltage level \cite{drost2012ringfilter}. As shown in the lowermost description of Fig. \ref{fig:sro-integ}, the $\phi$-to-$\phi$ transfer function is described by $K_\text{SRO}/s$ where $K_\text{SRO}$ is the switching gain of a \tip{SRO}.

\begin{figure}[t]
    \begin{center}
    \includegraphics[width=\columnwidth]{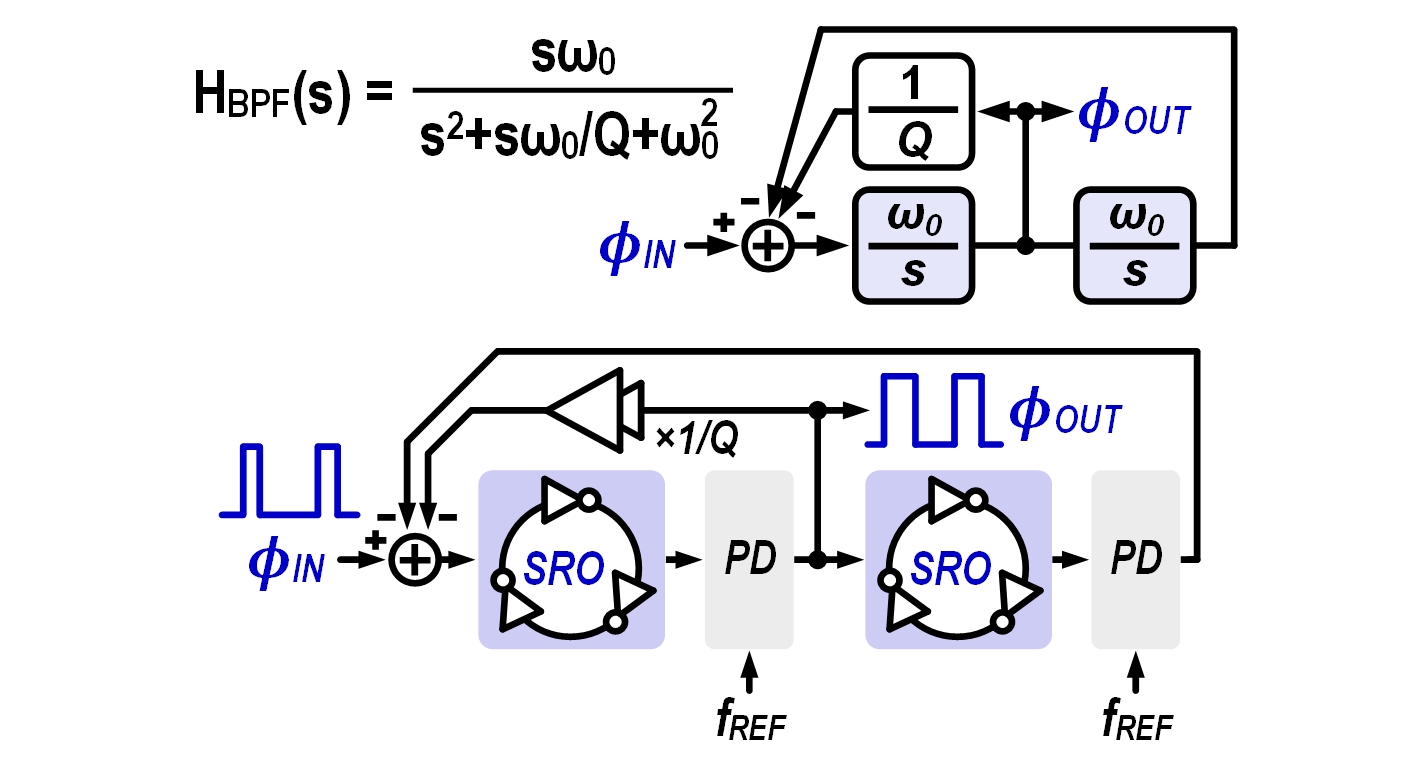}
    \caption{Block diagram of a time-domain second-order \tip{BPF} using \tip{SRO} as a core building block with a half-circuit representation.}\label{fig:bpf-concept}
    \end{center}
    \vspace{-5mm}
\end{figure}

Fig. \ref{fig:bpf-concept} shows a conceptual diagram for the implementation of a time-domain $\phi$-to-$\phi$ \tip{BPF}. It adopts the two-integrator-loop Tow-Thomas biquad topology \cite{tow1968biquad,thomas1971biquad} using \tip{SRO} as a lossless integrator ($\omega_{0}/s$). The \tip{PD} extracts the phase difference between two input signals and outputs the phase difference in a \tip{PWM} signal. The output of \tip{PD} is used to close the feedback loops where the inner feedback loop ensures the desired $Q$ factor and the outer feedback loop generates the high-pass shape of the \tip{BPF}. Overall, the time-domain \tip{BPF} receives the \tip{PWM} input and generates the \tip{PWM} output. Note that an external clock $f_\text{REF}$ is fed to the \tip{PD} in Fig. \ref{fig:bpf-concept} since it is represented as a simplified half-circuit diagram. If the two \tips{BPF} are placed in parallel to work as a pseudo-differential configuration as shown in Fig. \ref{fig:sro-integ}, the external clock $f_\text{REF}$ is no longer needed and the \tip{BPF} operates in a fully asynchronous way. The same consideration for a pseudo-differential topology also applies to the \tip{VTC} design as described in Section \ref{sec_vtc}.

\begin{figure}[t]
    \begin{center}
    \includegraphics[width=\columnwidth]{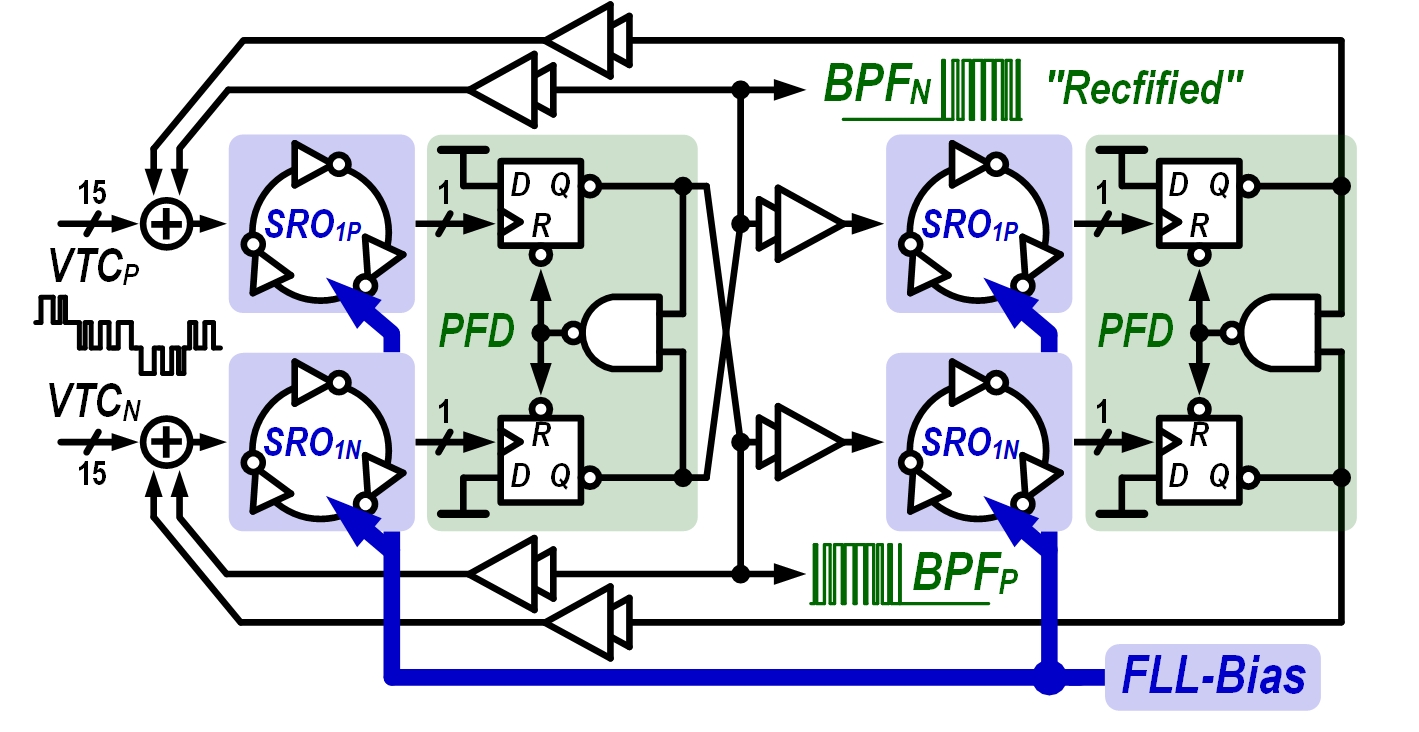}
    \caption{0.5 V supply time-domain rectifying \tip{BPF}.}\label{fig:bpf}
    \end{center}
    \vspace{-5mm}
\end{figure}

\begin{figure*}[t]
    \begin{center}
    \includegraphics[width=\textwidth]{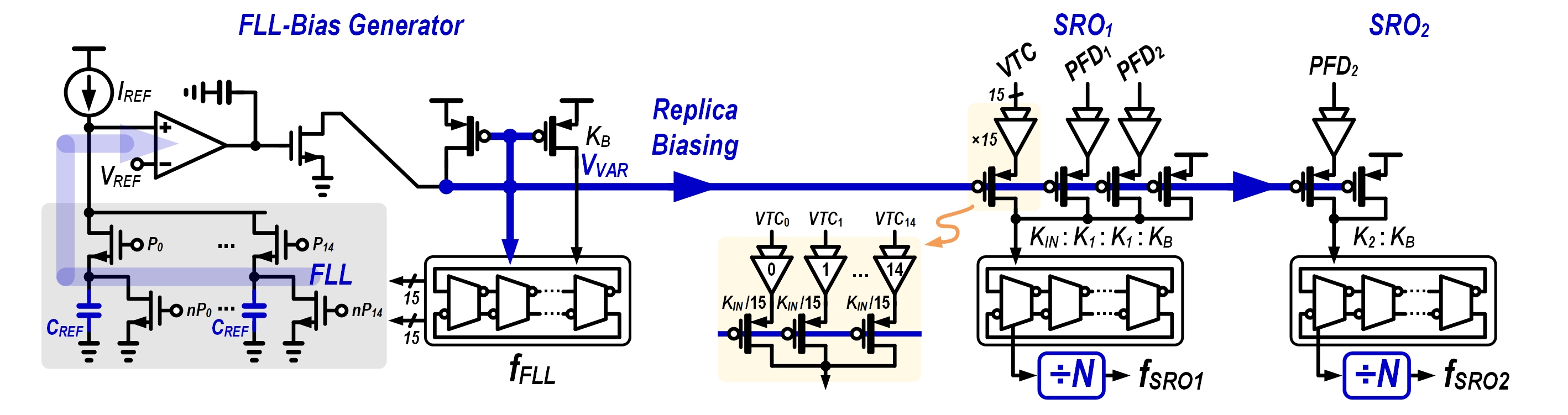}
    \caption{Schematic of \tip{FLL}-based bias generator and \tip{SRO}.}\label{fig:fll-bias}
    \end{center}
    \vspace{-5mm}
\end{figure*}

Fig. \ref{fig:bpf} shows the block diagram of the proposed time-domain \tip{BPF}. It receives the multi-phase \tip{PWM} output of the \tip{VTC} as an input signal and does not require an external clock. It incorporates four \tips{SRO} and two \tips{PFD}. The outputs of the \tip{BPF} are two single-phase \tip{PWM} signals. The two \tips{PFD} implemented in the \tip{BPF} offers an inherent rectification function in time-domain, which will be discussed in Section \ref{sec_rectifier}. A local \tip{FLL}-based bias generator provides the required bias voltage which are shared over the four \tips{SRO}. This bias voltage is different over 16-channel \tip{BPF} bank to set different center frequencies. Note that the outputs of first \tip{PFD} are crossed and connected to the \tips{SRO} with opposite polarities, to realize a subtraction function.

\begin{figure}[t]
    \begin{center}
    \includegraphics[width=\columnwidth]{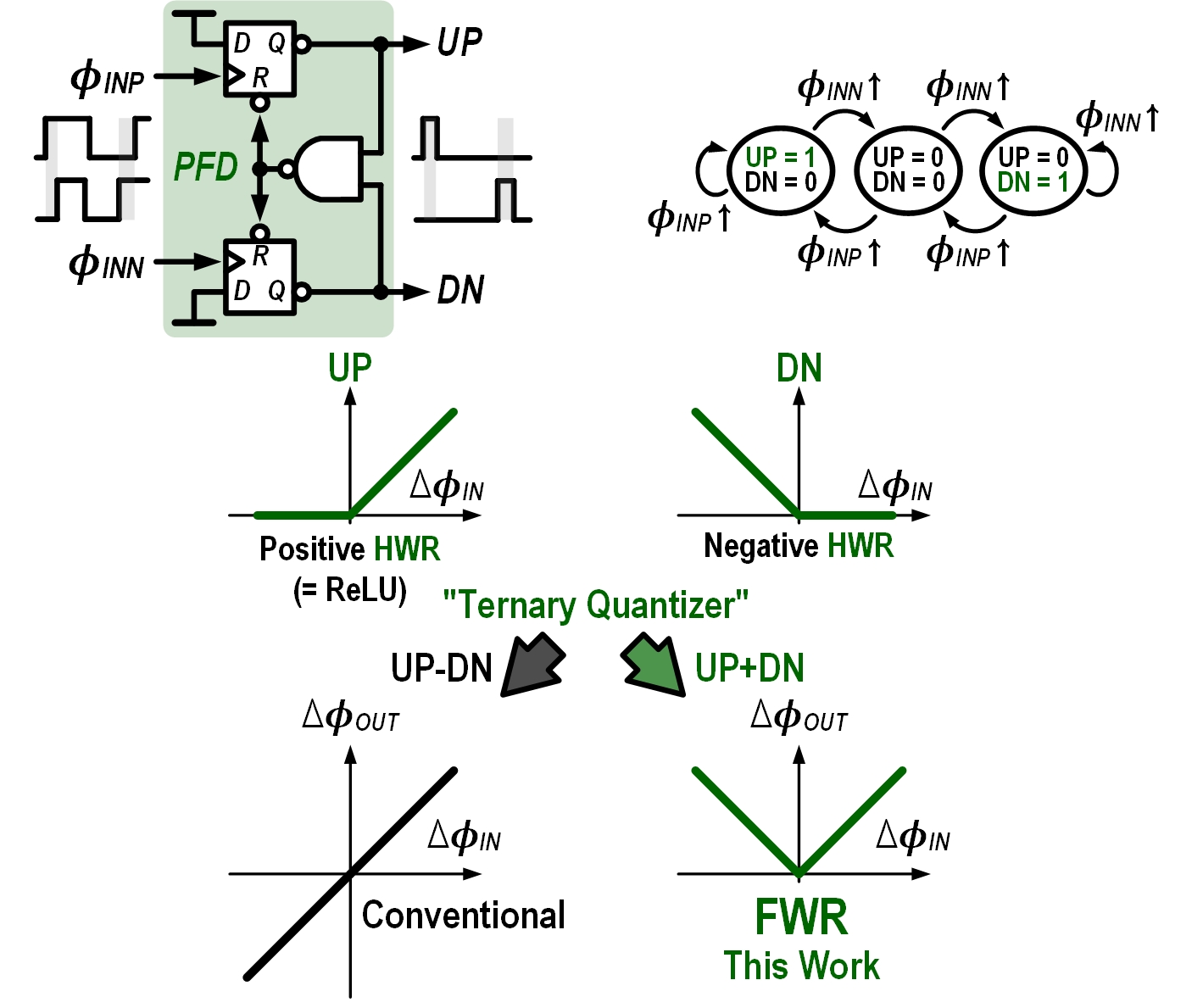}
    \caption{Proposed time-domain \tip{FWR} with operational principle of the \tip{PFD}.}\label{fig:fwr-pfd}
    \end{center}
    \vspace{-5mm}
\end{figure}

Fig. \ref{fig:fll-bias} shows a schematic of the \tip{FLL}-based bias generator and \tips{SRO} used in our \tip{BPF} design. The \tip{SRO} receives time-domain \tip{PWM} signals as input, such as \tip{VTC} and \tip{PFD}. All the \tip{PWM} signals are summed at the internal node of the \tip{SRO} and these signals drive the buffers which act as an array of current-mode \tips{DAC}. Therefore, the output frequency of \tip{SRO} is proportional to the sum of incoming \tip{PWM} signals. To realize different switching gains of \tip{PWM} inputs, switching transistors are differently sized. The unit current for current-\tip{DAC} operation is provided by a local \tip{FLL} circuit. The \tip{FLL} acts as a bias generator for realization of per-channel center frequency designs in \tips{BPF}, using a replica biasing scheme. As shown in Fig. \ref{fig:fll-bias}, the bias voltage $V_\text{VAR}$ is generated from a diode-connected pFET in the FLL-bias circuit. Therefore, the current-\tip{DAC} in \tip{SRO}\textsubscript{1-2} operates as a current mirror when $V_\text{VAR}$ is shared over the four \tips{SRO} from the \tip{FLL}-bias circuit. This means that the switching gain of each \tip{PWM} input signal is determined by $f_\text{FLL}$ and the sizing ratio of the current-mode \tip{DAC}. For example, the switching gain of \tip{VTC}-port is $K_\text{IN}f_\text{FLL}$ and the switching gain of \tip{PFD}\textsubscript{2}-port fed into the \tip{SRO}\textsubscript{2} is $K_\text{2}f_\text{FLL}$. Note that we adopt the same circuit structure from the sub-\tip{OSC} circuit in Section \ref{sec_vtc} which allows the \tip{BPF} to work at 0.5 V low-supply voltage. As discussed in \eqref{eq:fll}, the locking frequency of the \tip{FLL} circuit is proportional to $1/C_\text{REF}$. To cover the target range of \tip{BPF} center frequencies ranging from 100 Hz to 8 kHz in our design, a coarse-fine approach is used. The output of \tip{SRO} is divided coarsely by $N$ times using \tips{D-FF} and the $C_\text{REF}$ of \tip{FLL} circuit is fine controlled through proper sizing. The complete transfer function $H_\text{BPF}(s)$ of the proposed time-domain \tip{BPF} is given in \eqref{eq:bpf}. Its center frequency $\omega_{0}$ and $Q$-factor are given in \eqref{eq:bpf-w-Q} where the $Q$-factor is designed as 2 for each \tip{BPF} channel by proper sizing of the switching transistors in the \tips{SRO}. The stability of the proposed second-order time-domain \tip{BPF} is verified with a transient simulation.

\begin{equation} \label{eq:bpf}
    H_\text{BPF}(s)=\frac{\displaystyle \frac{sK_\text{IN}f_\text{FLL}K_\text{PFD}}{N}}
    {\displaystyle s^{2}+s\frac{K_{1}f_\text{FLL}K_\text{PFD}}{N}
    +\frac{K_{1}K_{1}f_\text{FLL}^{2}K_\text{PFD}^{2}}{N^{2}}}
\end{equation}
\begin{equation} \label{eq:bpf-w-Q}
    \omega_{0}=f_\text{FLL}K_\text{PFD}\sqrt{\frac{K_{1}K_{2}}{N}}\qquad Q=\sqrt{\frac{K_{2}}{K_{1}}}
\end{equation}

\subsection{Time-Domain Rectifier}\label{sec_rectifier}

\begin{figure*}[t]
    \begin{center}
    \includegraphics[width=\textwidth]{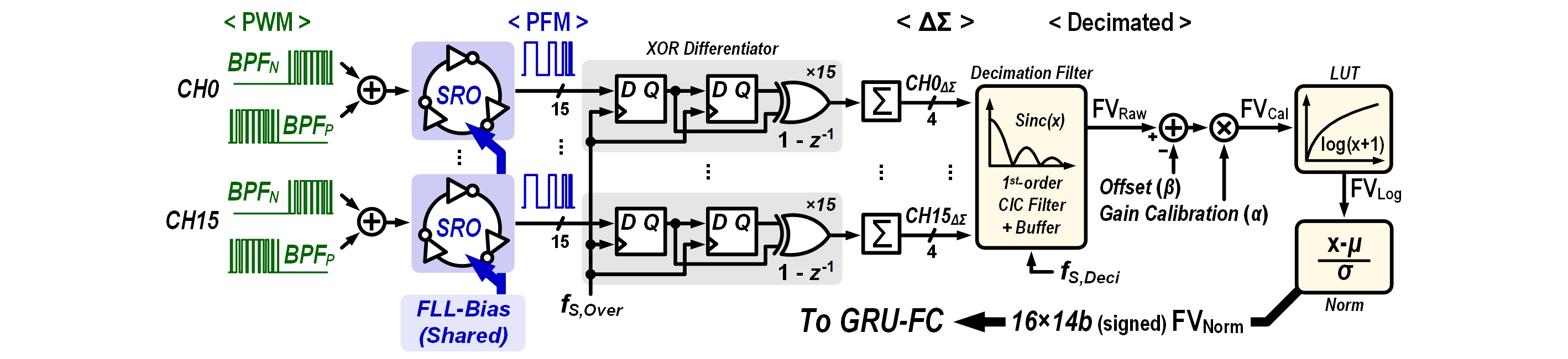}
    \caption{\tip{SRO}-based \tip{PFM} encoder, XOR differentiator, and subsequent post-processing blocks.}\label{fig:pfm-tdc}
    \end{center}
    \vspace{-5mm}
\end{figure*}

Fig. \ref{fig:fwr-pfd} shows the schematic of \tip{FWR}. The proposed time-domain \tip{FWR} is based on a simple \tip{PFD} circuit consisting of only two \tips{D-FF} and one NAND gate. Compared to the prior voltage-domain design \cite{yang2019vad} that required several scaling-unfriendly \tips{OTA}, references, and passive elements, this work offers an alternative solution that is fully compatible with standard logic gates. Fig. \ref{fig:fwr-pfd} shows a state diagram and an input-output characteristic of the \tip{PFD}. The \tip{PFD} circuit extracts the input phase difference $\Delta\phi_\text{IN}$, but at the same time asynchronously quantizes the phase difference using a ternary code with UP and down (DN) signals. As shown in the state diagram, there are three states that are activated by the rising edges of incoming \tip{PWM} signals ($\phi_\text{INP}$, $\phi_\text{INN}$). When both UP and DN signals are high, the NAND gate resets two \tips{D-FF} immediately, thereby making itself as a ternary quantizer. Since the state of \tip{PFD} stays the same unless a new rising edge arrives, the UP and DN signal represents a positively and negatively Half-Wave Rectified (HWR) phase difference $\phi_\text{IN}$, respectively. Interestingly, the UP signal has the same form as the \tip{ReLU} activation function widely used in modern \tips{DNN}. In conventional usage of such signals like in \tip{PLL} designs, they are subtracted to derive a linearized phase difference extractor. However, if we add them, a time-domain \tip{FWR} can be implemented. The \tip{PFD}-based \tip{FWR} benefits from its fully time-domain nature, that is, it does not exhibit a headroom-related saturation, assuming that the input signal swing ($\Delta\phi_\text{IN}$) is within $\pm 2\pi$ range. As shown in Fig. \ref{fig:bpf}, the proposed time-domain \tip{FWR} is seamlessly integrated within the time-domain \tip{BPF} circuit and thus the \tip{BPF} provides an inherent rectification function. The rectified \tip{PWM} signals ($BPF_\text{P/N}$) are summed at the subsequent \tip{PFM} stage as described in Fig. \ref{fig:pfm-tdc}.

\begin{figure}[t]
    \begin{center}
    \includegraphics[width=\columnwidth]{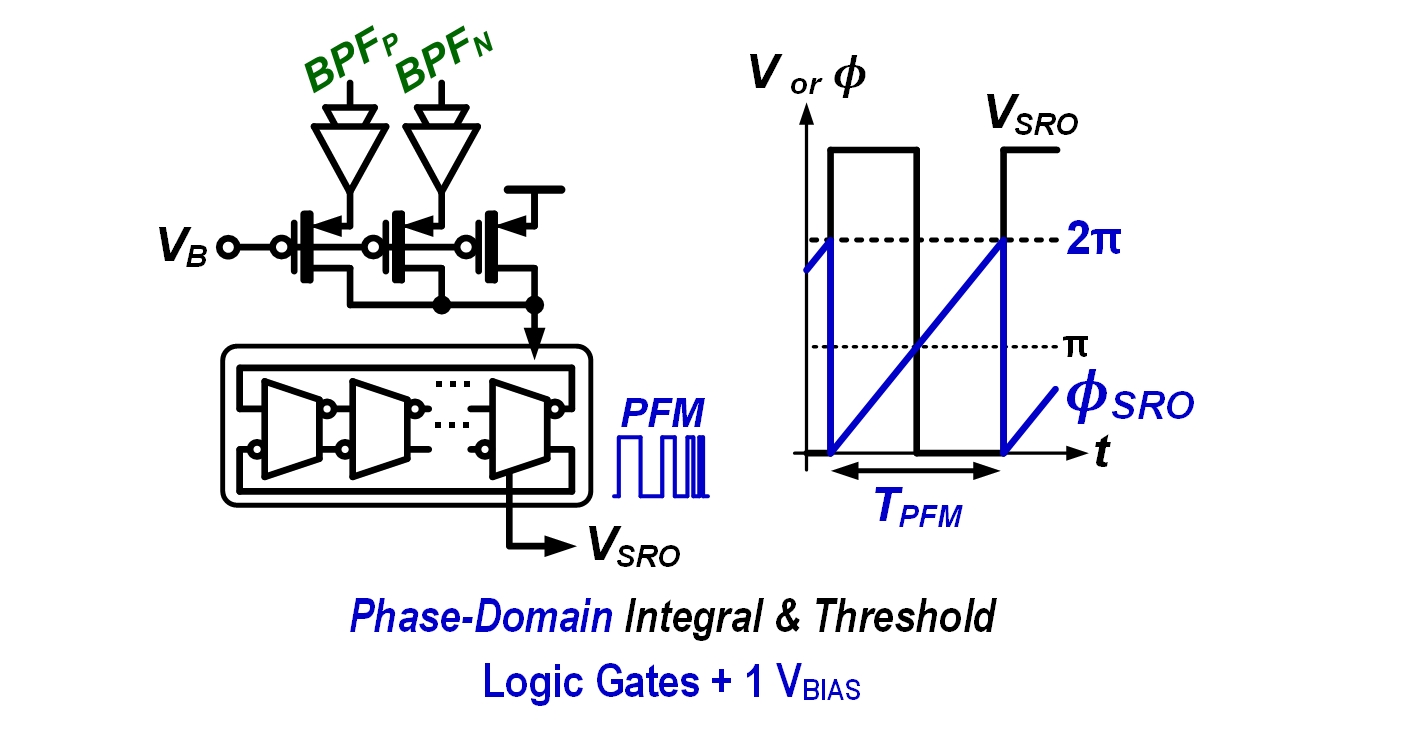}
    \caption{Proposed \tip{SRO}-based \tip{PFM} encoder.}\label{fig:pfm}
    \end{center}
    \vspace{-5mm}
\end{figure}

\subsection{Pulse-Frequency Encoder and Time-to-Digital Converter}\label{sec_tdc}

The analog \tip{FEx} designs presented in \cite{yang2019vad, wang2021kws} used an \tip{IAF} circuit which was originally proposed in \cite{mead1989a-vlsi}. The circuit converts an input current into a rate-encoded spiking \tip{PFM} signal and its spiking frequency is proportional to the input current magnitude. The \tip{IAF} circuit can be interpreted as a \tip{CCO} where the core oscillator topology is equivalent to a relaxation oscillator \cite{abidi1983relaxation}. However, the \tip{IAF} circuit is scaling-unfriendly because of its voltage-domain integral operation and voltage-domain static amplifier as discussed in Section \ref{sec_intro}. In this work, we propose to use the \tip{SRO} as a \tip{PFM} encoder instead of the \tip{IAF} circuit. As shown in Fig. \ref{fig:pfm}, the \tip{SRO} exploits an inherent phase-domain $2\pi$ threshold and its integral operation occurs in the phase-domain which is free from headroom issue. The proposed \tip{SRO}-based design offers a scaling-friendly implementation using only logic gates and bias generator without a static amplifier or passive components.

In previous designs \cite{yang2019vad, wang2021kws}, an asynchronous ripple-carry counter associated with a multi-bit register was used to quantize and sample the input \tip{PFM} signal. Interestingly, given the aforementioned interpretation of \tip{IAF} circuit as an oscillator, the signal flow from an \tip{IAF} to a counter builds an \tip{VCO}/\tip{CCO}-based $\Delta\Sigma$ modulator \cite{iwata1999vcoadc, elshazly2014sro}. However, the design approaches used in \cite{yang2019vad, wang2021kws} had two major problems. First, the ripple-carry counter exhibits metastability-induced data corruption when the sampling occurs at the instant of multibit transition of binary codes. Second, the output digital data from the asynchronous counter, which is $\Delta\Sigma$ modulated, was directly fed to the \tip{DNN} classifier without filtering of high-pass-shaped quantization noise. Our approach uses arrayed 1-bit XOR differentiators \cite{straayer2008vcoadc} to solve the metastability problem and an oversampling associated with a decimation filter to filter out quantization noise.

Fig. \ref{fig:pfm-tdc} shows the overview of implemented 16-channel \tip{SRO}-based \tip{PFM} encoder, XOR differentiator, and subsequent post-processing stages, also with the signal flow domains at the top. The \tip{SRO} receives rectified \tip{PWM} signals from the preceding \tip{BPF} stage and converts it into 15-phase \tip{PFM} signals. The same design of \tip{FLL} circuit discussed in Section~\ref{sec_bpf} is reused for biasing of the \tip{SRO} where the generated bias voltage is shared over 16 channels. As the ring-\tip{OSC} output is represented in thermometer-code, the XOR differentiator ensures the worst-case error to be within 1-\tip{LSB}. In addition, this 1-\tip{LSB} error is noise-shaped \cite{kim2010vcoadc} which can be eliminated through oversampling and decimation filtering. The thermometer-coded output data is aggregated to be represented as binary format and then filtered and decimated through a 1\textsuperscript{st}-order \tip{CIC} filter. We use 2\textsuperscript{10} decimation size, i.e., $f_\text{S,Deci}=f_\text{S,Over}/2^{10}$, and $f_\text{S,Deci}$ is used in the post-processing blocks which incorporates a programmable offset ($\beta$) subtractor to remove the DC offset due to a free-running component of the \tip{SRO}-based \tip{PFM} encoder. A programmable per-channel gain calibrator ($\alpha$) is used to correct inter-channel gain deviations caused by mismatch of SROs in the PFM encoder. A logarithmic compression using a LUT and a programmable input normalizer helps to increase the classification accuracy of the following \tip{GRU}-\tip{FC} neural network. The post-processing stage is clocked at 61\,Hz $f_\text{S,Deci}$ and thus its power dissipation is negligible.

\begin{figure}[t]
    \begin{center}
    \includegraphics[width=\columnwidth]{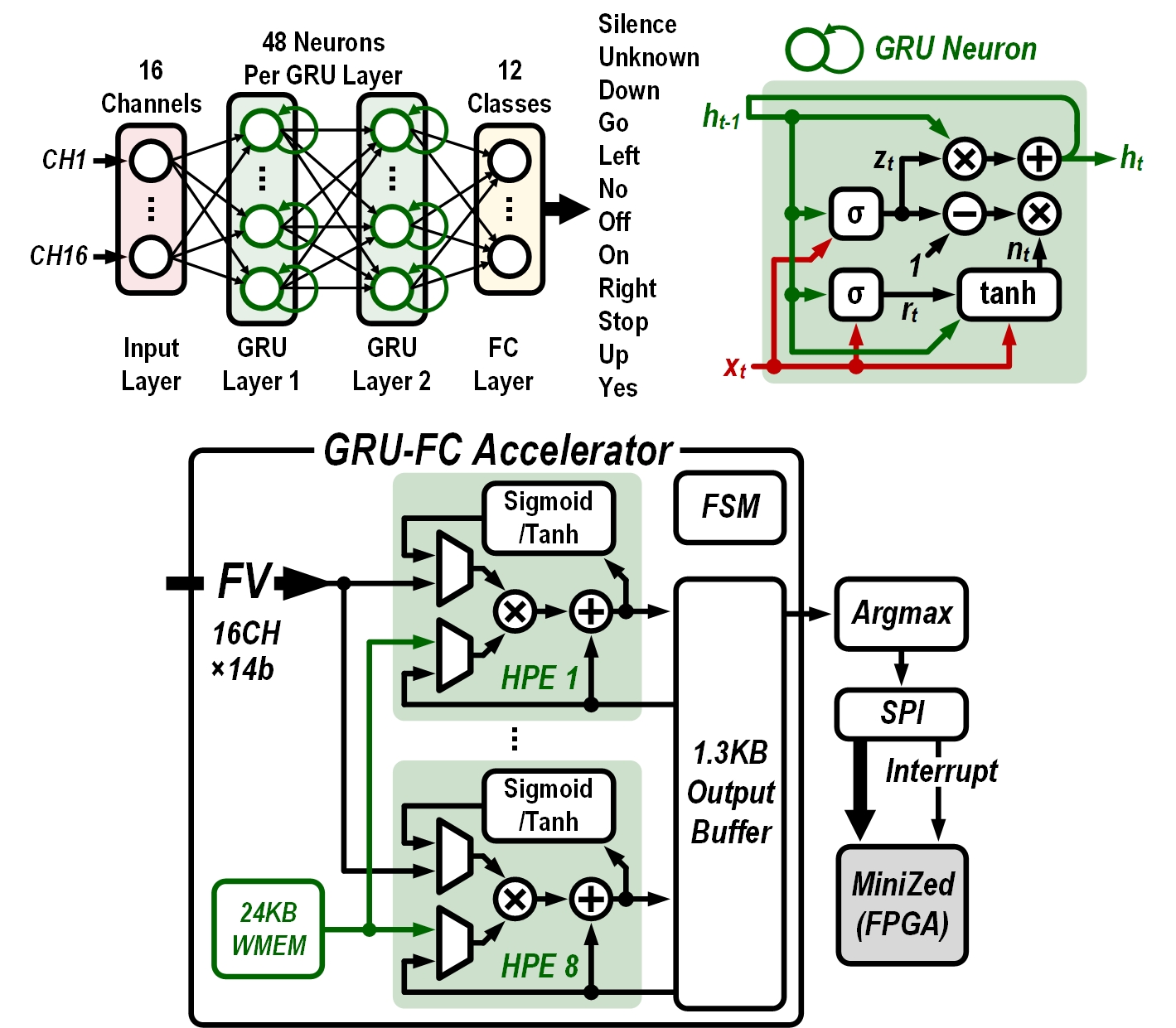}
    \caption{Architecture of the \tip{GRU}-\tip{FC} classification network (upper) and accelerator (lower).}\label{fig:gru}
    \end{center}
    \vspace{-5mm}
\end{figure}

\subsection{Recurrent Neural Network Accelerator}\label{sec_rnn}

Fig.~\ref{fig:gru} shows the architectures of the \tip{GRU}-\tip{FC} network and accelerator. The network has two \tip{GRU} layers with 48 units per layer and a final \tip{FC} layer that generates the confidence scores of the 12 classes. The network model size is entirely buffered within the 24\,KB \tip{WMEM}. The accelerator computes the \tip{KWS} classifier network and its input comes from the normalizer shown in Fig. \ref{fig:pfm-tdc}. The \tip{WMEM} block is implemented by on-chip \tip{SRAM} compiled based on the foundry-provided 6-transistor (6T) bit-cell. The classifier weights are loaded into \tip{WMEM} over the SPI interface. The accelerator has 8 Heterogeneous Processing Elements (HPEs) controlled by a \tip{FSM}. Each HPE has a 14-bit multiplier, a 24-bit accumulator, and a LUT-based Sigmoid/Tanh unit. Partial sums of the multiply-accumulate operations and outputs are stored in a shared 1.3\,KB \tip{SRAM} output buffer. Multiplexers before each multiplier operand select inputs from the normalizer, the Sigmoid/Tanh unit, the \tip{WMEM}, and the output buffer to compute element-wise vector multiplication/addition and hyperbolic functions in the \tip{GRU} \tip{RNN}. The high $V_{TH}$ device library is used for logic synthesis to reduce leakage current. Output scores of the classifier are fed to the argmax decoder, which outputs the class with the highest score. The classification result is transmitted over the SPI interface with an interrupt flag to the external host, which is a MiniZed board with a Xilinx Zynq-7007S SoC.

\subsection{Network Training}\label{sec_train}

\begin{figure}[t]
    \begin{center}
    \includegraphics[width=\columnwidth]{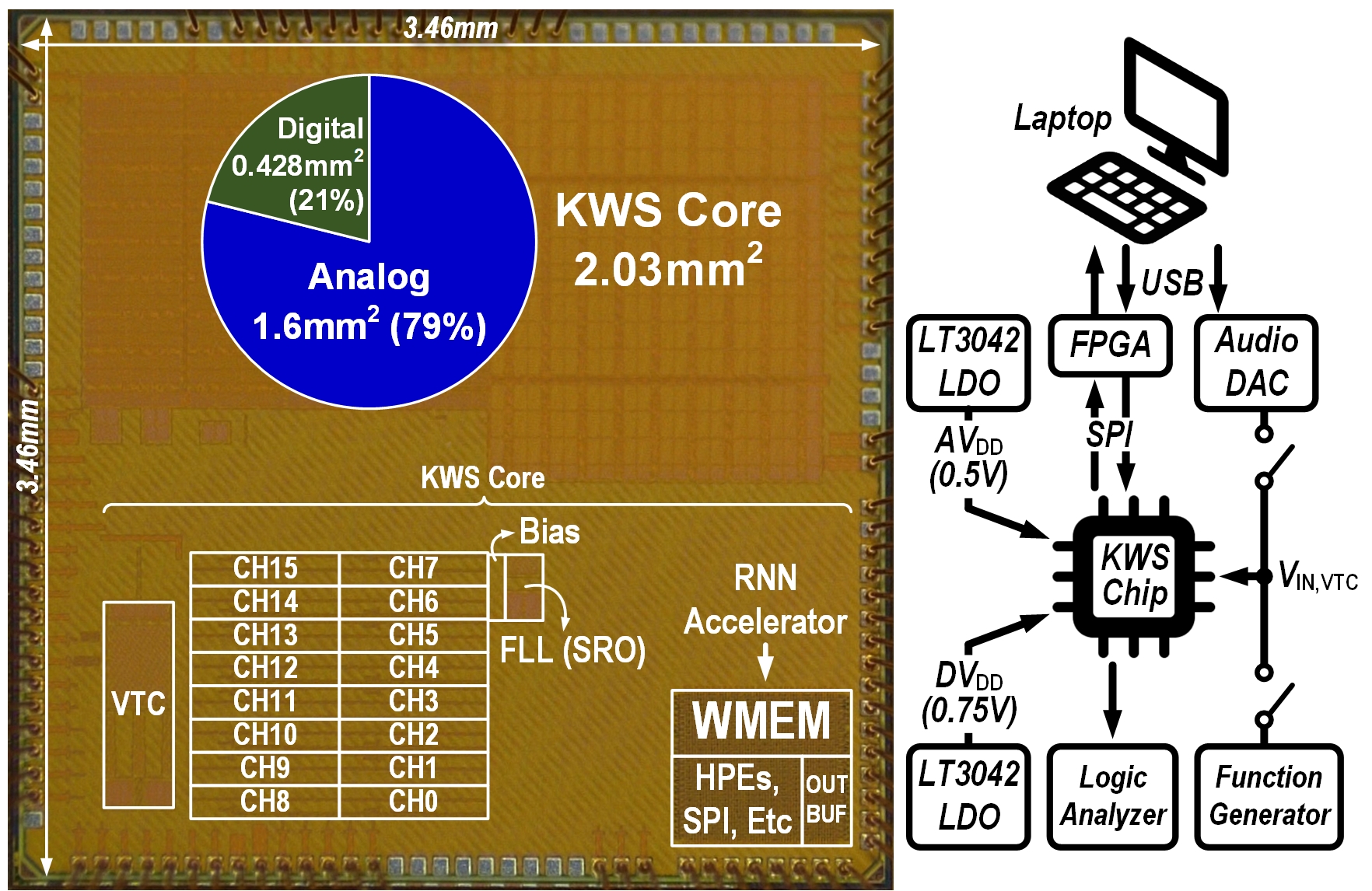}
    \caption{Chip photograph of the prototyped KWS IC with a block-diagram of the measurement setup.}\label{fig:chip_photo}
    \end{center}
    \vspace{-5mm}
\end{figure}

\paragraph{Dataset preparation}{Our \tip{GSCD} training set is composed of 38,463 samples. The number of samples in the ``Silence" class is 4,044 which are randomly sampled from the background noise tracks in the dataset. The ``Unknown" class also has 4,044 samples which are randomly chosen words outside the target 12 classes. As for the test set, we used the standard \tip{GSCD} test set\footnote{\url{http://download.tensorflow.org/data/speech_commands_test_set_v0.02.tar.gz}} which has roughly equal number of samples (around 400) among the 12 target classes. Thus, the ratio between training and test set is around 8:1.} As shown in the measurement setup of Fig.~\ref{fig:chip_photo}, the samples from our entire training and test set were played from a laptop to $V_\text{IN,VTC}$ through a USB sound card \tip{DAC} (Sound Blaster E1). We normalized the \tip{GSCD} samples with the mean and standard deviation of the entire samples such that the amplitude of $V_\text{IN,VTC}$ is set to $\sim$250\,mV\textsubscript{PP}. The corresponding $\text{FV}_\text{Raw}$ from all samples were recorded. They were then corrected for the DC offset ($\beta$) and the inter-channel gain deviation ($\alpha$). After applying the logarithmic compression, we then normalize the $\text{FV}_\text{Raw}$ with the mean ($\mu$) and standard deviation ($\sigma$) of the recorded feature vectors from the entire \tip{GSCD} training set. This resulting vector called $\text{FV}_\text{Norm}$ (see Fig.~\ref{fig:overall}) are then presented as inputs to the \tip{GRU}-\tip{FC} classifier during training. The same $\mu$ and $\sigma$ are applied to $\text{FV}_\text{Log}$ of the test set to generate the corresponding $\text{FV}_\text{Norm}$ for \tip{KWS} evaluation.
\paragraph{Training schedule} The network is built in the PyTorch 1.8 framework and trained for 200 epochs using the AdamW optimizer \cite{loshchilov2018adamw} with an initial learning rate of 1e-3 and 0.01 weight decay. The ReduceLROnPlateau learning rate scheduler is used with a decay factor of 0.8 and patience of 3 epochs. The lowest learning rate is 5e-4. Using quantization-aware training, the activations and weights are quantized to 14-bits and 8-bits respectively.

\begin{figure}[t]
    \begin{center}
    \includegraphics[width=\columnwidth]{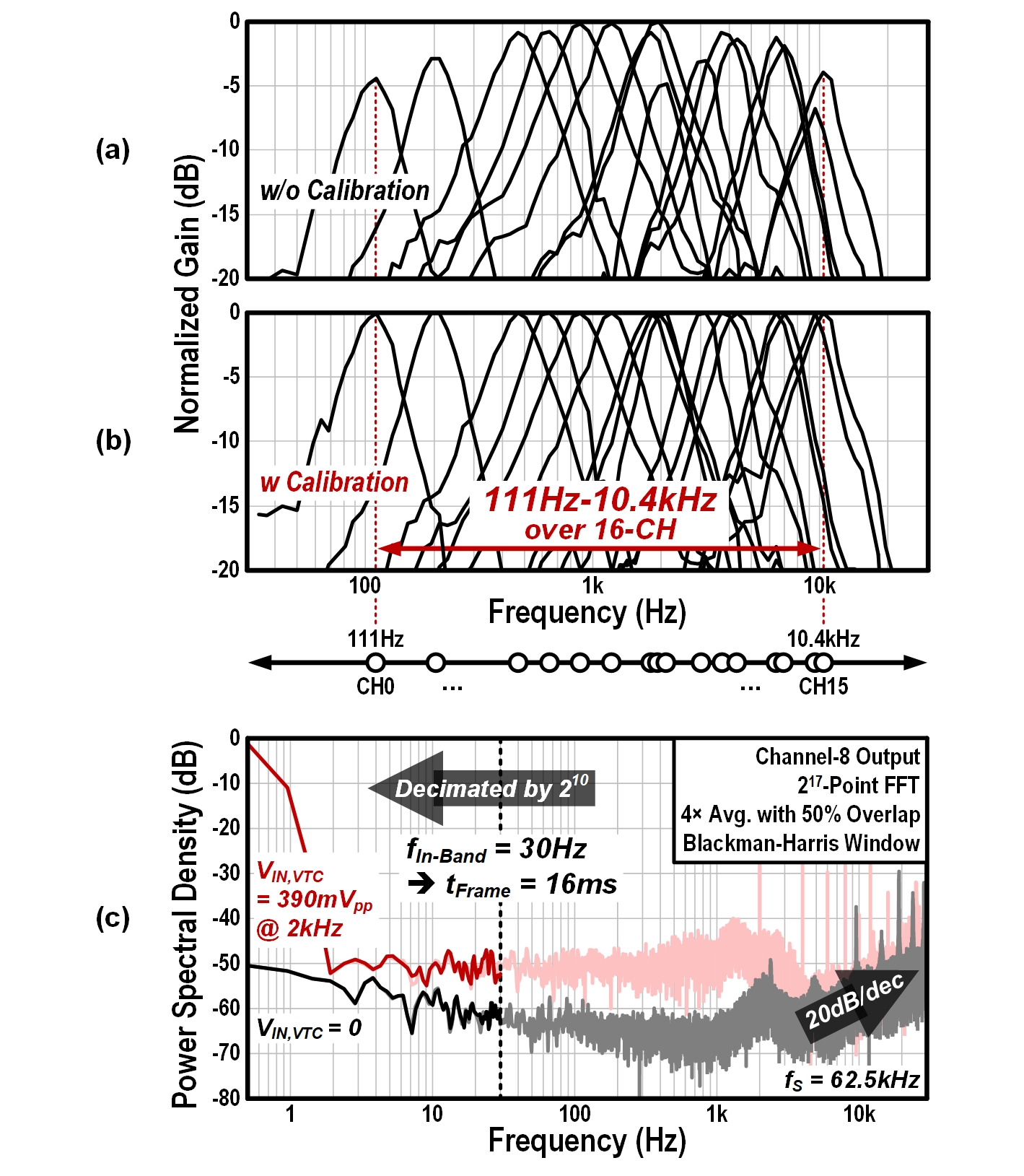}
    \caption{Measured frequency response of the \tip{FEx} (a) without per-channel correction, (b) with per-channel gain ($\alpha$) correction, and (c) output spectrum of $\text{FV}_\text{Raw}$ (after decimation filter) in channel 8 as shown in Fig.~\ref{fig:pfm-tdc}.}\label{fig:meas_bpf-spectrum}
    \end{center}
    \vspace{-5mm}
\end{figure}

\section{Measurement Results}\label{sec_measurement}

Fig.~\ref{fig:chip_photo} shows the \tip{KWS} IC, which is fabricated in TSMC 65 nm CMOS LP process with an active area of 2.03mm\textsuperscript{2} for the \tip{KWS} core. The area occupied by the analog and digital circuits is 1.6mm\textsuperscript{2} (79\%) and 0.428mm\textsuperscript{2} (21\%) respectively in the \tip{KWS} core. The \tip{GRU}-\tip{FC} neural network accelerator and the associated peripherals in the digital circuits are synthesized from a standard auto place-and-route (P\&R) flow.

Fig. \ref{fig:meas_bpf-spectrum}(a) and (b) show the measured frequency response of the 16-channel \tip{FEx} with and without per-channel gain calibration. In this case, $V_\text{IN,VTC}$ was connected to a function generator. The center frequencies of the 16 \tip{BPF} channels range from 111\,Hz to 10.4\,kHz. The center frequencies are distributed according to the Mel scale therefore low-frequency ($<$1 kHz) channels are spaced further apart than high-frequency channels. As shown in Fig. \ref{fig:meas_bpf-spectrum}(a), the measured gain curve before the calibration shows the inter-channel gain deviations which are caused by systematic mismatches from the \tip{SRO}-based \tip{PFM} encoder. The main cause of gain deviation is the voltage bias ($V_\text{VAR}$ in Fig. \ref{fig:fll-bias}) which is generated from a single \tip{FLL} circuit and it is shared over the 16-channel \tip{SRO} as depicted as `\tip{FLL} (\tip{SRO})' in the chip photograph. We expect this systematic mismatch due to distribution of the voltage bias can be improved with a better layout floorplan, for example with a centralized placement of the bias circuits while the random mismatch can be improved with larger sizing of the biasing transistors.

Fig. \ref{fig:meas_bpf-spectrum}(c) shows the measured output spectrum of channel 8 for two different input conditions of the \tip{VTC}; the black curve is obtained with a zero input condition while the red curve is obtained with a 2 kHz sinusoidal input of 390 mV\textsubscript{PP}. Here, the amplitude of the input to the \tip{VTC} circuit was assumed to be sufficiently large, and our future work will include an additional ultra-low-power pre-amplifier \cite{Han2013amp} before the \tip{VTC} circuit. The oversampling clock frequency that is fed into the XOR differentiator is 62.5\,kHz. It is clearly seen that the output spectrum has a first-order noise-shaping property with a 20\,dB/dec slope for both input conditions. After the feature data is decimated by 2\textsuperscript{10}, the in-band frequency is limited as 30\,Hz which is translated into a 16\,ms frame shift or 61\,frame/s throughput, and so the 16-channel \tip{FV} is generated every 16\,ms. The integrated in-band noise with zero input is calculated as 248 $\mu$V\textsubscript{RMS}, which is dominated by $1/f$ noise. When the input amplitude is increased to 390 mV\textsubscript{PP}, the in-band noise is dominated by thermal noise. We believe that the noise increase is caused by a higher running frequency of the \tip{SRO}, since the phase noise of ring oscillators increases with operating frequency \cite{abidi2006phasenoise}.

\begin{figure}[t]
    \begin{center}
    \includegraphics[width=\columnwidth]{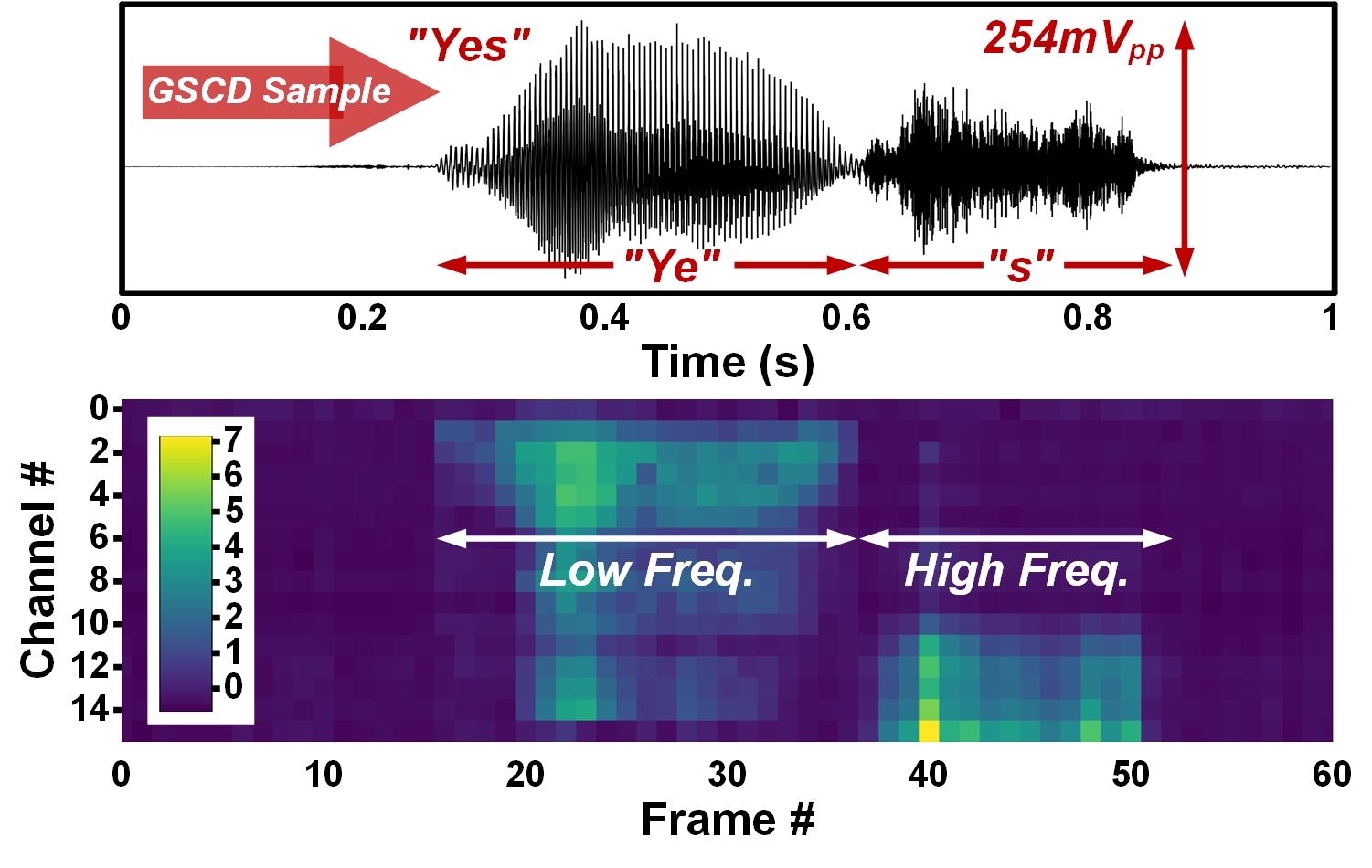}
    \caption{Measured audio response of the \tip{FEx} with an applied sample keyword from \tip{GSCD}.}\label{fig:meas_audio}
    \end{center}
    \vspace{-5mm}
\end{figure}

\begin{figure}[t]
    \begin{center}
    \includegraphics[width=\columnwidth]{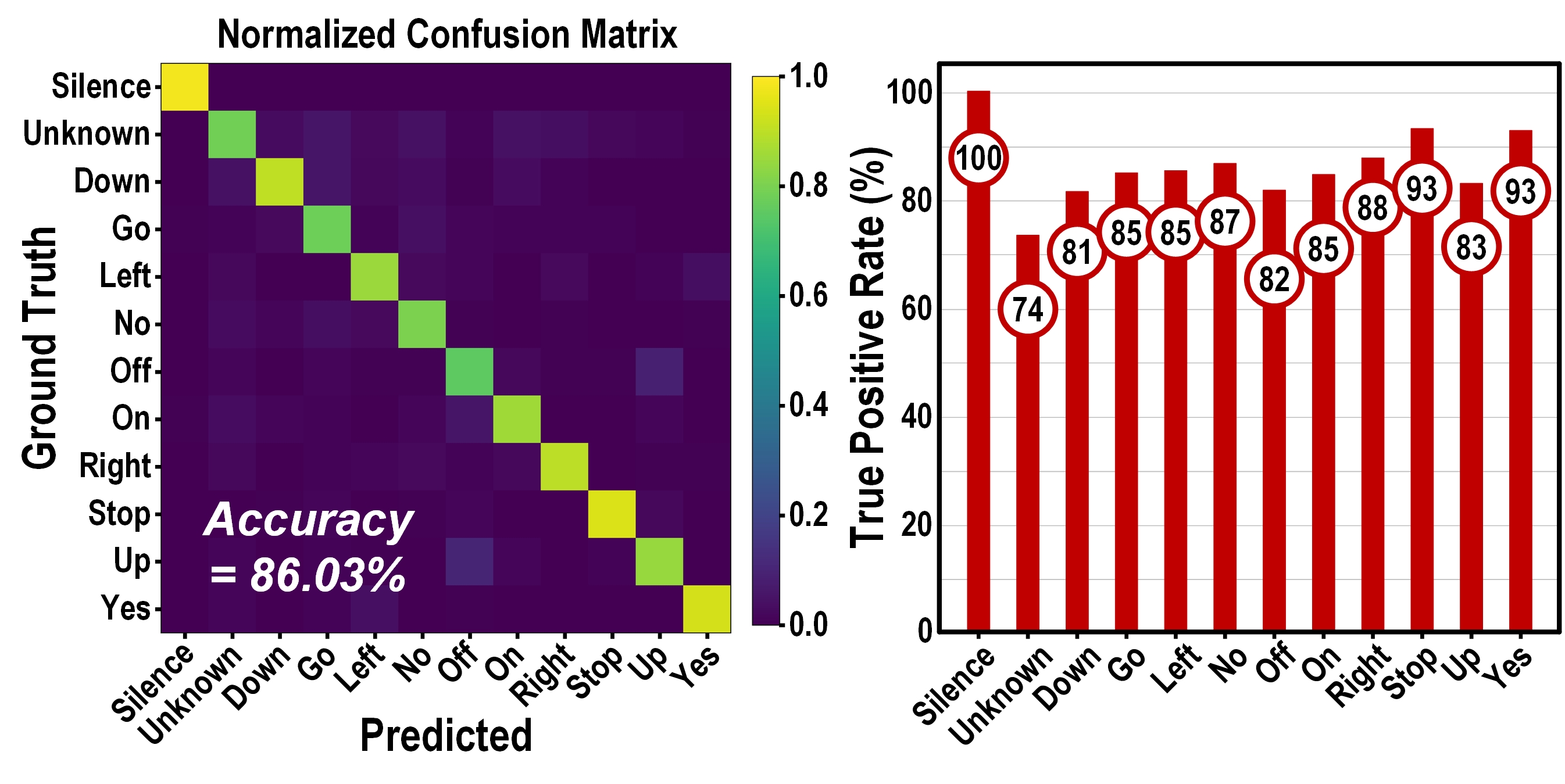}
    \caption{Measured \tip{KWS} accuracy on the \tip{GSCD} test set. A confusion matrix (left) where the magnitudes are normalized between 0 and 1, and a plot of the true positive rates over 12 different classes (right) are shown.}\label{fig:meas_accuracy}
    \end{center}
    \vspace{-5mm}
\end{figure}

\begin{figure}[t]
    \begin{center}
    \includegraphics[width=\columnwidth]{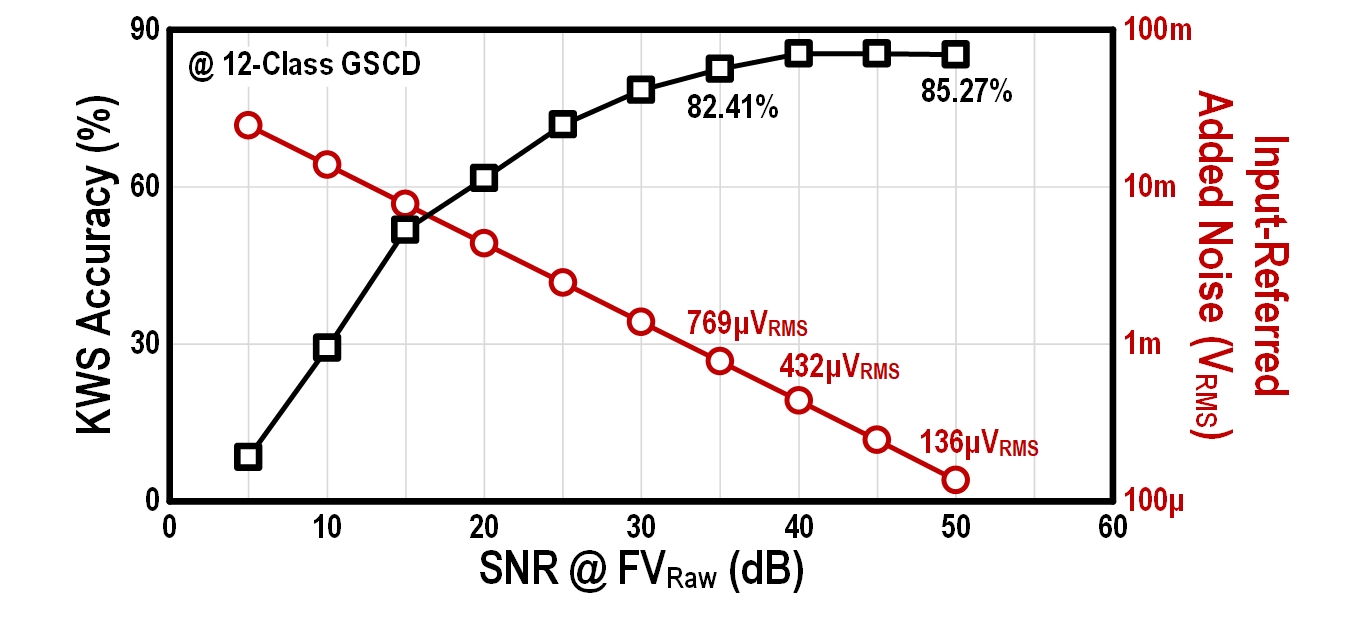}
    \caption{\tip{KWS} accuracy obtained over different \tip{SNR} levels.}\label{fig:meas_accuracy_SNR}
    \end{center}
    \vspace{-5mm}
\end{figure}

\begin{figure}[t]
    \begin{center}
    \includegraphics[width=\columnwidth]{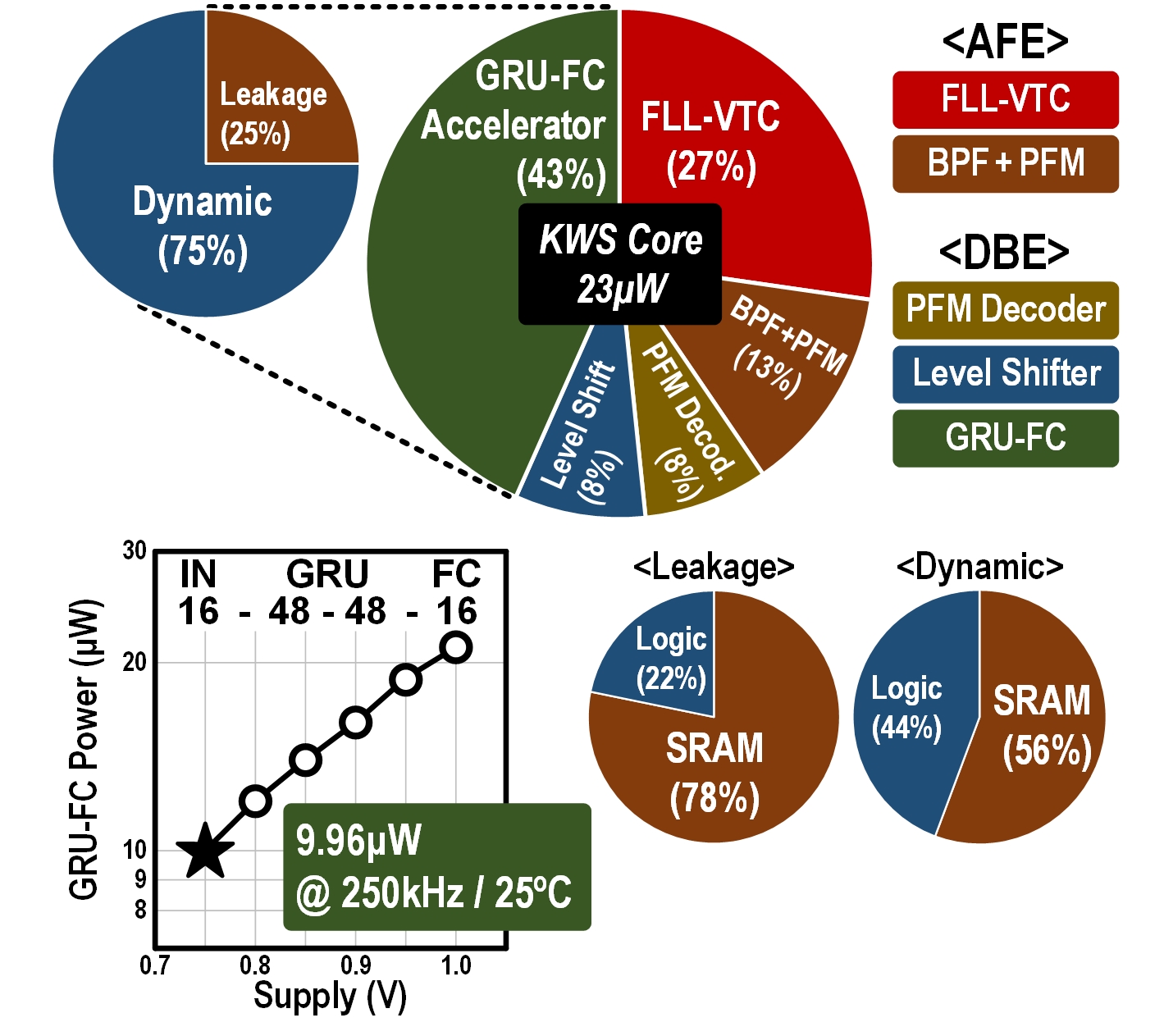}
    \caption{Power breakdown of the \tip{KWS} core implemented in the proposed IC (see Fig.~\ref{fig:overall}).}\label{fig:meas_powerbreak}
    \end{center}
    \vspace{-5mm}
\end{figure}

Fig. \ref{fig:meas_audio} shows the measured audio response of the \tip{FEx}. A 254\,mV\textsubscript{PP} ``Yes'' keyword sample from \tip{GSCD} is selected and applied to the \tip{VTC} while measuring the \tip{FEx} output. The magnitudes of \tip{FV} in this figure are normalized by subtracting DC offset and dividing by the standard deviation of the sample clip, for better visualization. It is clearly seen that the 16-channel \tip{FV} has higher response at low frequencies for the ``Ye" sound, and at higher frequencies for the ``s'' sound.

Fig. \ref{fig:meas_accuracy} shows the measured \tip{KWS} accuracy of the prototype chip obtained using the full 12-class verification flow of the \tip{GSCD}v2 \cite{warden2018gscd}. The 12 classes include ``Silence", ``Unknown", and 10 target keywords. As shown in the measurement setup in Fig.\,\,\ref{fig:chip_photo}, the generated \tips{FV} ($\text{FV}_\text{Raw}$ in Fig.~\ref{fig:overall}) from our time-domain \tip{FEx} are recorded using the \tip{GSCD} training set that is fed to the \tip{VTC} of our chip ($V_\text{IN,VTC}$ in Fig.~\ref{fig:chip_photo}), to train the classifier network. The 16 ms frame window and 16 ms frame shift (stride) are used for recording, so there is no overlap between two consecutive frames. The detected class is the most active output at the end of the \tip{GSCD} sample. The prototype \tip{KWS} IC achieves an overall 86.03\% accuracy with the \tip{GSCD} test set. The measured true positive rates show that ``Silence" is the easiest class with 100\% accuracy and the classifier performed the best on two keywords, ``Stop" and ``Yes," with 93\% accuracy. The most challenging class is ``Unknown" class since it includes 25 non-target keywords such as ``Happy" and ``Dog," which requires the classifier to train more parameters with larger model capacity to improve the accuracy. We expect that the detection accuracy of ``Unknown" and thus the overall accuracy on this \tip{KWS} dataset will improve with a larger network model, but at the expense of additional power consumption and silicon area.
The state-of-the-art accuracy on \tip{GSCD} using \tip{GRU}-\tips{RNN} is 94.2\% \cite{Zhang2017helloedge} with a network which has 499\,KB parameters and running on a Cortex-M7 microcontroller. This network size would require 21$\times$ more on-chip memory leading to higher power consumption and chip area. 


\begin{table*}[ht]
\centering
\caption{Performance Comparison Table - Analog \tip{FEx}}\label{comp_table_FEx}
\begin{tabular}{||c||c|c|c|c|c|}
\hline
{Analog FEx} &
\begin{tabular}[c]{@{}c@{}}
{M. Yang}\\{JSSC 2016 \cite{yang2016coch}}
\end{tabular} &
\begin{tabular}[c]{@{}c@{}}
{K. Badami}\\{JSSC 2016 \cite{badami2016vad}}
\end{tabular} &
\begin{tabular}[c]{@{}c@{}}
{M. Yang}\\{JSSC 2019 \cite{yang2019vad}}
\end{tabular} &
\begin{tabular}[c]{@{}c@{}}
{S. Oh}\\{JSSC 2019 \cite{oh2019vad}}
\end{tabular} &
\textbf{This Work}
\\ \hline
{Process (nm)} &
180 &
90 &
180 &
180 &
\textbf{65}
\\ \hline
{Area/Ch. (mm\textsuperscript{2})} &
0.26 &
0.13 &
0.1 &
- &
\textbf{0.1}
\\ \hline
{Architecture} &
g\textsubscript{m}C-\tip{BPF} &
g\textsubscript{m}C-\tip{BPF} &
g\textsubscript{m}C-\tip{BPF} &
Mixer &
\textbf{OSC-BPF}
\\ \hline
{Number of Ch.} &
64$\times$2 &
16 &
16 &
32{\color{Maroon}\textsuperscript{A}} &
\textbf{16}
\\ \hline
{Freq. Range (Hz)} &
8-20k &
75-5k &
100-5k &
75-4k &
\textbf{111-10.4k}
\\ \hline
{Supply (V)} &
0.5 &
- &
0.6 &
1.4 &
\textbf{0.5}
\\ \hline
{Power ($\mu$W)} &
55 &
6 &
0.38 &
0.06 &
\textbf{9.3}
\\ \hline
{Frame Shift (ms)} &
- &
31.25{\color{Maroon}\textsuperscript{B}} &
10 &
512{\color{Maroon}\textsuperscript{A}} &
\textbf{16}
\\ \hline
{Dynamic Range (dB)} &
55 &
45 &
40{\color{Maroon}\textsuperscript{C}} &
47 &
\textbf{54.89}
\\ \hline
{FoM\textsubscript{S,DR} (dB)} &
- &
82.3 &
91.5 &
91.33 &
\textbf{93.11}
\\ \hline
{Target Task} &
General Purpose &
\multicolumn{3}{c|}{
VAD
} &
\textbf{KWS}
\\ \hline
{Building Blocks} &
\begin{tabular}[c]{@{}c@{}}
BPF, ADM\\
\end{tabular} &
\begin{tabular}[c]{@{}c@{}}
LNA, BPF\\FWR, LPF
\end{tabular} &
\begin{tabular}[c]{@{}c@{}}
LNA, BPF\\FWR, IAF
\end{tabular} &
\begin{tabular}[c]{@{}c@{}}
LNA, Mixer\\LPF, DSP
\end{tabular} &
\begin{tabular}[c]{@{}c@{}}
\textbf{VTC}\\\textbf{Rec-BPF}\\\textbf{PFM}
\end{tabular}
\\ \hline
{Support SE Mic} &
No &
\checkmark &
No &
\checkmark &
\textbf{\checkmark}
\\ \hline
{Parallel FEx} &
\checkmark &
\checkmark &
\checkmark &
No &
\textbf{\checkmark}
\\ \hline
\multicolumn{6}{|l|}{
{\color{Maroon}\textsuperscript{A}}With 32-32-16-2 FC neural network $\qquad$
{\color{Maroon}\textsuperscript{B}}f\textsubscript{LPF} = 16 Hz
}\\
\multicolumn{6}{|l|}{
{\color{Maroon}\textsuperscript{C}}Measured by the firing rate range of the IAF only, excluding output noise of the feature vector
}\\ \hline
\end{tabular}
\begin{equation}
\label{eq:Norm_Power}
    \mathrm{P_{Norm}}=
    \frac{P(1-r)}{1-r^{n}}\cdot
    \frac{20\mathrm{k}}{f_{H}}
    \qquad\qquad
    r=\left(\frac{f_{L}}{f_{H}}\right)^{1/(n-1)}
\end{equation}
\begin{equation}
\label{eq:FoM}
    \mathrm{FoM_{S,DR}}=
    \mathrm{DR}+
    10\cdot\mathrm{log}_{10}
    \left(\frac{1}{\mathrm{P_{Norm}}\cdot 2\cdot \mathrm{FrameShift}}\right)
\end{equation}
\end{table*}


\begin{table*}[ht]
\centering
\caption{Performance Comparison Table - \tip{KWS}}\label{comp_table_KWS}
\begin{tabular}{||c||c|c|c|c|c|c|c|}
\hline
{KWS} &
\begin{tabular}[c]{@{}c@{}}
{S. Zheng}\\{TCAS-I 2019 \cite{zheng2019kws}}
\end{tabular} &
\begin{tabular}[c]{@{}c@{}}
{H. Dbouk}\\{JSSC 2021 \cite{dbouk2021kws}}
\end{tabular} &
\multicolumn{2}{c|}{
\begin{tabular}[c]{@{}c@{}}
{W. Shan}\\{JSSC 2021 \cite{shan2021kws}}
\end{tabular}} &
\begin{tabular}[c]{@{}c@{}}
{J. Giraldo}\\{VLSI 2019 \cite{giraldo2019kws}}
\end{tabular} &
\begin{tabular}[c]{@{}c@{}}
{D. Wang}\\{ISSCC 2021 \cite{wang2021kws}}
\end{tabular} &
\textbf{This Work}
\\ \hline
{-} &
\multicolumn{4}{c|}{
Off-Chip ADC
} &
On-Chip ADC &
\multicolumn{2}{c|}{
On-Chip Analog FEx
}
\\ \hline
{Process (nm)} &
28 &
65 &
\multicolumn{2}{c|}{
28
} &
65 &
65 &
\textbf{65}
\\ \hline
{Area (mm\textsuperscript{2})} &
1.29{\color{Maroon}\textsuperscript{A}} &
4.13{\color{Maroon}\textsuperscript{A}} &
\multicolumn{2}{c|}{
0.23{\color{Maroon}\textsuperscript{A}}
} &
1.52 &
2.71{\color{Maroon}\textsuperscript{B}} &
\textbf{2.03}
\\ \hline
{SRAM (KB)} &
52 &
38 &
\multicolumn{2}{c|}{
2
} &
32 &
20 &
\textbf{27}
\\ \hline
{Clock (Hz)} &
2.5M &
1G &
\multicolumn{2}{c|}{
40k
} &
250k &
120k &
\textbf{250k}
\\ \hline
{FEx} &
Digital &
- &
\multicolumn{2}{c|}{
Digital
} &
Digital &
Analog Voltage &
\textbf{Analog Time}
\\ \hline
{Classifier} &
CNN &
RNN &
\multicolumn{2}{c|}{
CNN
} &
RNN &
SNN (MLP) &
\textbf{RNN}
\\ \hline
{KWS Power ($\mu$W)} &
141{\color{Maroon}\textsuperscript{A}} &
11000{\color{Maroon}\textsuperscript{A}} &
\multicolumn{2}{c|}{
0.51{\color{Maroon}\textsuperscript{A}}
} &
16.1 &
0.205-0.570 &
\textbf{23}
\\ \hline
Frame Shift (ms) &
10 &
20 &
\multicolumn{2}{c|}{16} &
16 &
100 &
\textbf{16}
\\ \hline
{Latency (ms)} &
10 &
0.04 &
\multicolumn{2}{c|}{
64
} &
16 &
100 &
\textbf{12.4}
\\ \hline
{Dataset} &
TIDIGITS &
\multicolumn{6}{c|}{
GSCD
}
\\ \hline
\begin{tabular}[c]{@{}c@{}}
{Number of Classes}\\{(Keywords)}
\end{tabular} &
2 &
\begin{tabular}[c]{@{}c@{}}
7{\color{Maroon}\textsuperscript{C}}\\(6)
\end{tabular} &
\begin{tabular}[c]{@{}c@{}}
2{\color{Maroon}\textsuperscript{C}}\\(1)
\end{tabular} &
\begin{tabular}[c]{@{}c@{}}
5{\color{Maroon}\textsuperscript{C}}\\(4)
\end{tabular} &
\begin{tabular}[c]{@{}c@{}}
12{\color{Maroon}\textsuperscript{D}}\\(10)
\end{tabular} &
\begin{tabular}[c]{@{}c@{}}
5{\color{Maroon}\textsuperscript{C}}\\(4)
\end{tabular} &
\begin{tabular}[c]{@{}c@{}}
\textbf{12{\color{Maroon}\textsuperscript{D}}}\\\textbf{(10)}
\end{tabular}
\\ \hline
\begin{tabular}[c]{@{}c@{}}
{Accuracy (\%)}
\end{tabular} &
96 &
90.38 &
\begin{tabular}[c]{@{}c@{}}
97.3{\color{Maroon}\textsuperscript{E}}
\end{tabular} &
\begin{tabular}[c]{@{}c@{}}
91.7{\color{Maroon}\textsuperscript{E}}
\end{tabular} &
90.87 &
90.2 &
\textbf{86.03}
\\ \hline
{Support SE Mic} &
\multicolumn{4}{c|}{
Off-Chip ADC
} &
No &
No &
\textbf{\checkmark}
\\ \hline
\multicolumn{8}{|l|}{
{\color{Maroon}\textsuperscript{A}}Excluding off-chip ADC $\qquad$
{\color{Maroon}\textsuperscript{B}}Including SNN chip \cite{wang2020snn} $\qquad$
{\color{Maroon}\textsuperscript{C}}Excluding ``Unknown" word detection as a distinct class}\\
\multicolumn{8}{|l|}{
{\color{Maroon}\textsuperscript{D}}2 non-keywords (Silience/Unknown) + 10 keywords $\quad$
{\color{Maroon}\textsuperscript{E}}Accuracy is reported from 16-bit GSCD samples; the design excludes the 16-bit ADC}
\\ \hline
\end{tabular}
\end{table*}

Fig.~\ref{fig:meas_accuracy_SNR} shows the dependence of the \tip{KWS} classification accuracy on added noise levels to the recorded feature vector $\text{FV}_\text{Raw}$ (see Fig.~\ref{fig:overall}). We first computed the average power $P_\text{Avg,GSCD}$ using the recorded $\text{FV}_\text{Raw}$ (see Section~\ref{sec_train}). Then Gaussian noise of different standard deviation values ($\sigma^{2}=P_\text{Avg,Noise}$) were added to create different $\text{SNR}$ values following the equation below.
\begin{align}
    \text{SNR}=10\log_{10}
    \left( \frac{P_\text{Avg,GSCD}}{P_\text{Avg,Noise}} \right)
    \label{eq:SNR}
\end{align}
The noise is randomly generated for each training epoch and test evaluation. For each \tip{SNR} case, our \tip{GRU}-\tip{FC} network is retrained using the noisy training set and the noisy test set is used to evaluate the classification accuracy. The proposed \tip{KWS} IC ensures $<$1\% accuracy drop even for noise levels up to 432\,$\mu$V\textsubscript{RMS} (input-referred to $V_\text{IN,VTC}$), or 40\,dB \tip{SNR}.

Fig.~\ref{fig:meas_powerbreak} shows the power breakdown of the \tip{KWS} IC. As it is stated in Section~\ref{sec_intro}, this paper focuses on the \tip{KWS} core only therefore the power breakdown in Fig.~\ref{fig:meas_powerbreak} does not include energy harvester, low-dropout regulator, and voltage reference circuits. The total power consumption of the \tip{KWS} core is 23\,$\mu$W when it is measured at 25$^{\circ}$C room temperature. The \tip{GRU}-\tip{FC} neural network accelerator accounts for 43\% of the \tip{KWS} core power. When the 16IN-48H-48H-12C \tip{GRU}-\tip{FC} network is updated at 250 kHz clock frequency and 0.75\,V supply voltage while performing continual inference on random \tip{GSCD} samples with 16 ms frame shift, the accelerator consumes 9.96\,$\mu$W. The accelerator power consumption can be further decomposed into dynamic power (75\%) and leakage (or static) power (25\%). The leakage power is dominated by the \tip{SRAM} block (78\%) while both logic (44\%) and \tip{SRAM} (56\%) contributed rather evenly to the dynamic power. We expect that leakage power can be reduced with custom memory cells \cite{shan2021kws}.

Table \ref{comp_table_FEx} compares the performance of our time-domain \tip{FEx} with the state-of-the-art voice processing analog \tip{FEx} \cite{yang2016coch, yang2019vad, badami2016vad, oh2019vad}. The proposed \tip{FEx} circuit is the first that demonstrated ring-oscillator-based \tip{BPF} topology used for the \tip{KWS} task. It supports \tip{SE} microphones, thereby offering a lower system-level power. Unlike sequential \tip{FEx}, our parallel \tip{FEx} does not lose frequency-selective information at any time \cite{oh2019vad}. To allow a fair comparison with previously reported designs with a variety of frame shifts, we derive a Schreier \tip{FoM} \cite{schreier2005dsm} \eqref{eq:FoM}, widely used for \tips{ADC}. The Schreier FoM considers the trade-off between \tip{DR} and bandwidth, also accounting for power consumption. For near DC input \tips{ADC}, the bandwidth is replaced with a reciprocal of conversion time \cite{chae2013dsm}. As a band-pass filtered signal is demodulated into baseband (DC) after the rectifier \cite{yang2019vad, badami2016vad} or mixer \cite{oh2019vad} stage, we consider analog \tip{FEx} as a DC-input \tip{ADC} with a pre-processing stage. The \tip{FoM} equation \eqref{eq:FoM} includes the normalized power consumption $P_\text{Norm}$ (\ref{eq:Norm_Power}) proposed in \cite{yang2016coch}, the \tip{DR}, and the frame shift. The frame shift is part of the denominator of \eqref{eq:FoM} because the \tips{FV} are generated in every frame shift. The amount of integrated in-band noise is reduced with a larger decimation window (i.e., averaged over the longer time interval) in our design and also in \cite{chae2013dsm} where the number of \tip{ADC} cycles was used for decimation window. The proposed \tip{FEx} records the best Schreier FoM among the state-of-the-art designs. In addition, the time-domain processing circuits offer better technology scaling, and will outperform voltage-domain designs \cite{yang2016coch, yang2019vad, badami2016vad, oh2019vad} in terms of power and area when implemented in advanced technology nodes.

Table \ref{comp_table_KWS} compares the performance of our KWS IC with other state-of-the-art \tip{KWS} ICs \cite{zheng2019kws, dbouk2021kws, shan2021kws, giraldo2019kws, wang2021kws}. This work uses an on-chip analog \tip{FEx} while other works needed an off-chip high-resolution (16-bit) \tip{ADC} \cite{zheng2019kws, shan2021kws}. Sometimes even the digital \tip{FEx} and \tip{ADC} were implemented off-chip \cite{dbouk2021kws}. Furthermore, only this work and \cite{giraldo2019kws} support the essential ``Unknown" class to be detected as a distinct class, which is the most challenging class in the \tip{GSCD} test set. As such, it implies that concessions would be made in terms of \tip{KWS} accuracy for \cite{dbouk2021kws, shan2021kws, wang2021kws}, or a larger model size will be required for the classifier to uphold the accuracy, leading to additional power and area costs. In addition, the proposed chip supports \tip{SE} microphone interface and the \tip{KWS} task is verified with a \tip{SE} input condition. This work shows competitive performance and better system-level power efficiency by using a low-power \tip{MEMS} \tip{SE} microphone instead of the differential microphone used in \cite{giraldo2019kws}. Last but not least, our prototype chip is the first silicon-verified analog \tip{FEx}-based voice processing IC that demonstrates 12-class \tip{KWS} task on \tip{GSCD}, using an on-chip classifier.

Our belief is that the 5\% degradation in the classification accuracy of our \tip{KWS} IC (86\%) as compared to the software model accuracy (91\%, Section~\ref{sec_software}) is mainly due to the increased noise floor when the input amplitude is high, as shown in Fig.~\ref{fig:meas_bpf-spectrum}(c). Advanced noise suppression techniques such as chopper stabilization \cite{enz1996chop} and dynamic element matching \cite{ha2019bioz} when applied to the front-end, will help mitigate the accuracy discrepancy. Our time-domain \tip{FEx} still needs to address the per-chip gain calibration requirement due to the mismatch of analog circuits, which is not necessary in a fully-digital approach \cite{giraldo2019kws}. For this, as discussed in the paragraph in Section~\ref{sec_measurement} describing Fig.~\ref{fig:meas_bpf-spectrum}(a), improved layout floorplan and larger device sizes accompanied with mismatch-aware \tip{DNN} training \cite{yang2019vad} will be another opportunity to remove the calibration requirement.

\section{Conclusion}\label{sec_conclusion}

We have presented a low-power \tip{KWS} chip that exploits ring-oscillator-based time-domain processing circuits. Implemented in a 65\,nm CMOS process, it consumes 23\,$\mu$W power dissipation with a power supply of 0.5\,V for analog circuits and 0.75\,V for digital circuits. The nested analog \tip{FLL} enhances the linearity of \tip{VTC}, and thus facilitates the use of single-ended microphones as discussed in Section \ref{sec_vtc}. The usage of \tip{PFD} as a time-domain \tip{FWR} shows significantly reduced implementation cost in comparison with a voltage-domain design. The \tip{PFM} functionality is realized using a \tip{SRO}, instead of the conventional \tip{IAF} circuit to obviate the need for scaling-unfriendly voltage-domain circuits. Table \ref{comp_table_FEx} shows that the proposed time-domain \tip{FEx} achieves the state-of-the-art \tip{DR}-based Schreier \tip{FoM}. The on-chip integrated \tip{GRU}-\tip{FC} digital back-end circuit processes incoming audio \tips{FV} with a 16\,ms frame shift using only $\sim$10\,$\mu$W power, demonstrating $>$86\% classification accuracy with only 12.4\,ms latency on the 12-class \tip{GSCD} \tip{KWS} task. We expect that the proposed time-domain processing techniques can be further expanded in other domains and thus provide various design opportunities for power-efficient circuits, such as the fully time-domain \tip{ReLU} activation unit shown in Fig. \ref{fig:fwr-pfd} for \tips{DNN}. The improvement directions as discussed in Section~\ref{sec_measurement} along with \tip{DR} enhancement techniques such as front-end automatic gain control will enable the time-domain \tip{FEx} to be applied to more challenging real-world audio-inference tasks. A 35-class \tip{KWS} on \tip{GSCD} \cite{rybakov2020kws} or a streaming-mode \tip{KWS} can be such examples.


\section*{Acknowledgment}
The authors would like to thank Frank K. Gürkaynak and Beat Muheim from ETH Zürich, for their technical
support of the digital circuits in this IC technology and Taekwang Jang from ETH Zürich for the valuable discussions on analog frequency-locked loop.


\bibliographystyle{IEEEtran}
\bibliography{myref}


\begin{IEEEbiography}
[{\includegraphics[width=1in,height=1.25in,clip,keepaspectratio]
{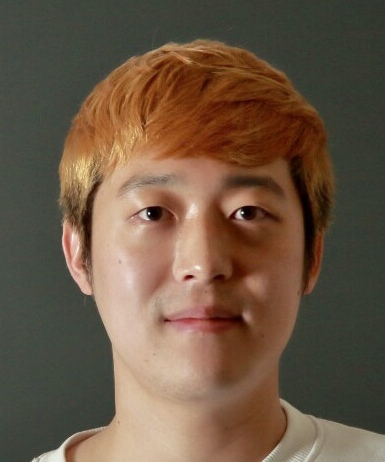}}]
{Kwantae Kim}
(Member, IEEE) received the B.S., M.S., and Ph.D. degrees in the School of Electrical Engineering, Korea Advanced Institute of Science and Technology (KAIST), Daejeon, South Korea, in 2015, 2017, and 2021, respectively. 

From 2015 to 2017, he was also with Healthrian R\&D Center, Daejeon, South Korea, where he designed bio-potential readout IC for mobile healthcare solutions. In 2020, he was a Visiting Student with the Institute of Neuroinformatics, University of Zürich and ETH Zürich, Zürich, Switzerland, where he is currently working as a Postdoctoral Researcher since 2021. His research interests include analog/mixed-signal ICs for time-domain processing, in-memory computing, bio-impedance sensor, and neuromorphic audio sensor.
\end{IEEEbiography}

\begin{IEEEbiography}
[{\includegraphics[width=1in,height=1.25in,clip,keepaspectratio]
{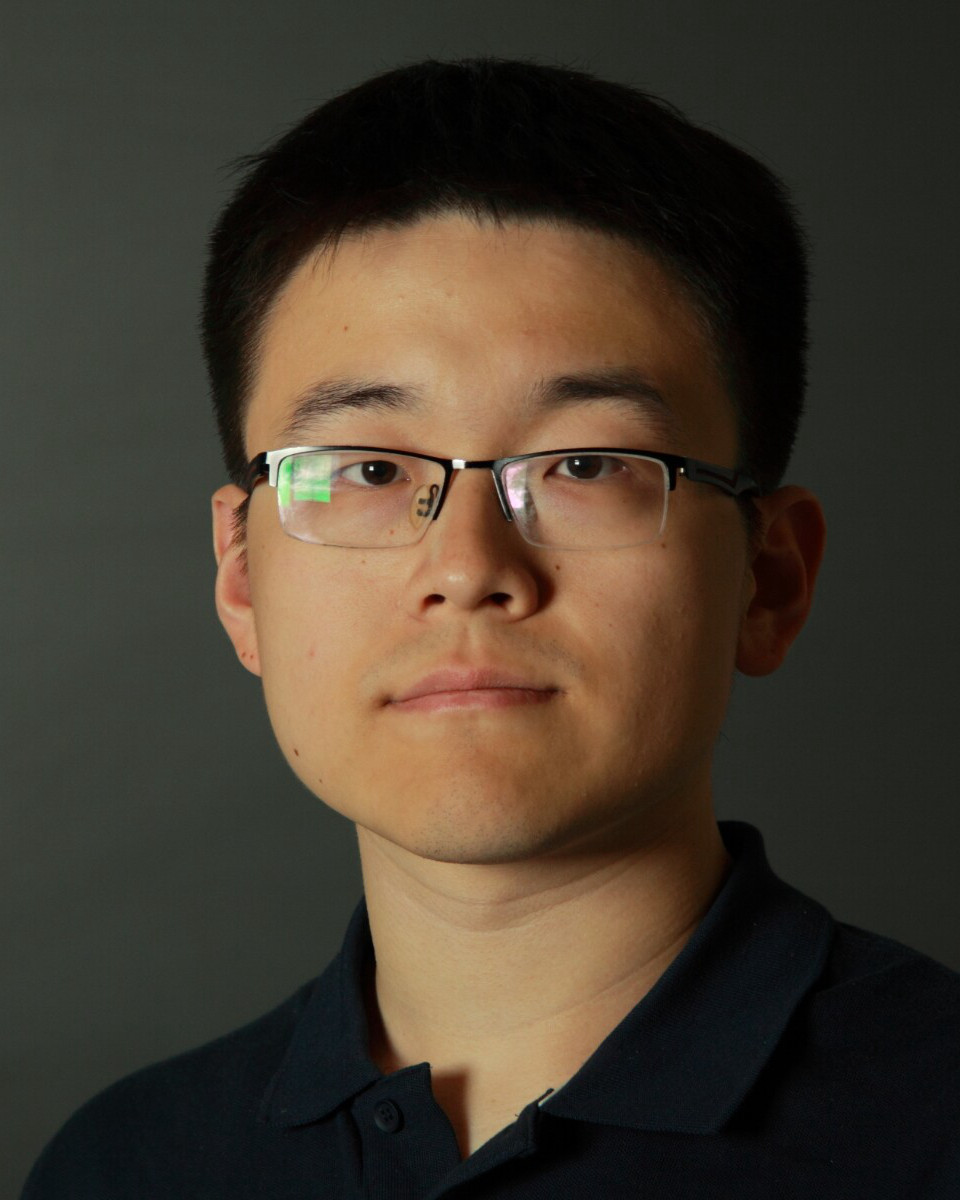}}]
{Chang Gao}
(Member, IEEE) received the B.S. degree in Electronics from University of Liverpool, Liverpool,
UK and Xi'an Jiaotong-Liverpool University, Suzhou, China, and the master’s degree in analog and digital integrated circuit design from Imperial College London, London, UK. He was awarded his Doctoral degree at the Institute of Neuroinformatics, University of Zurich and ETH Zurich, Zurich, Switzerland in Dec. 2021 and is joining TU Delft as an Assistant Professor in 2022.
His current research interests include computer architectures for deep learning with emphasis on recurrent neural networks.
\end{IEEEbiography}

\begin{IEEEbiography}
[{\includegraphics[width=1in,height=1.25in,clip,keepaspectratio]
{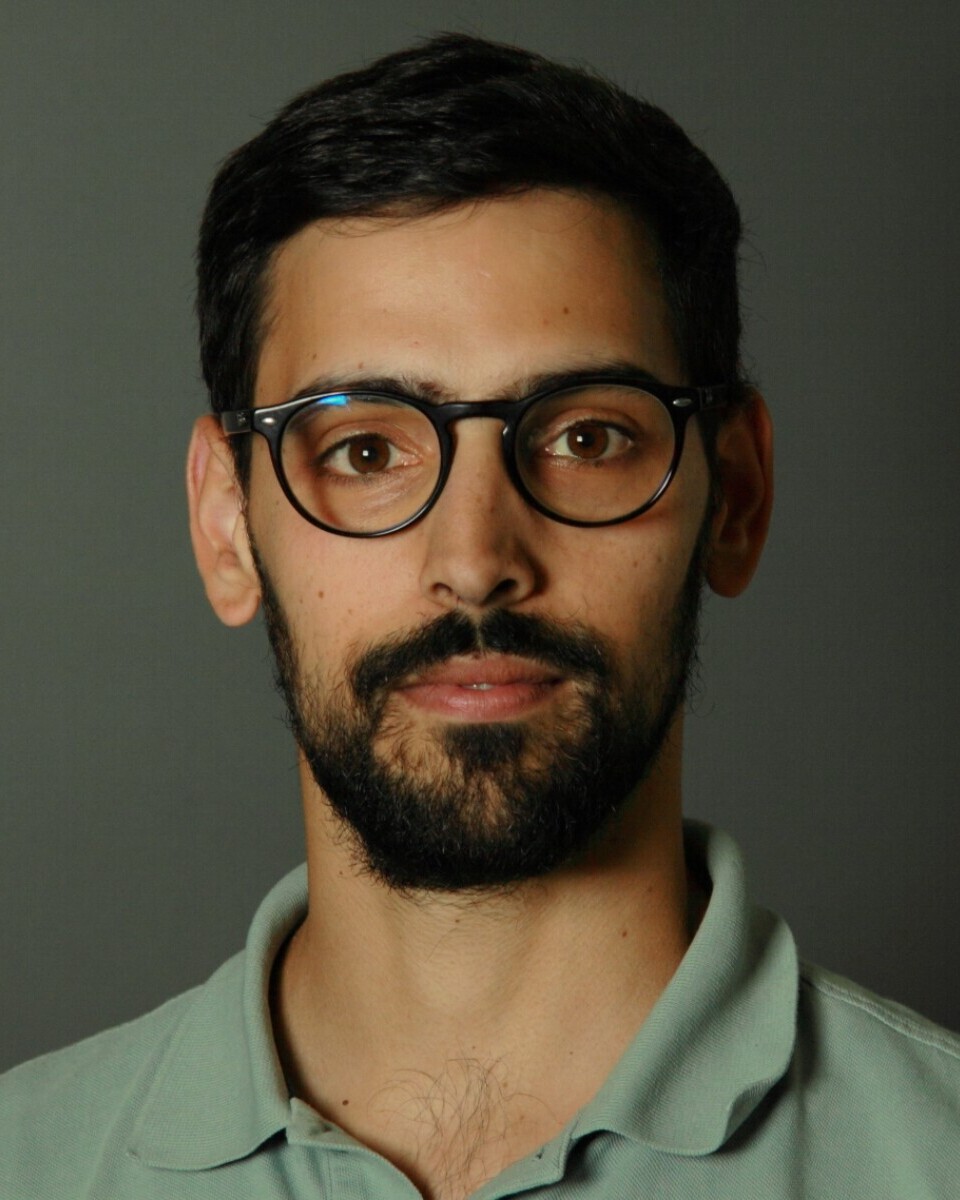}}]
{Rui Graça}
received the B.Sc. and M.Sc. degrees in Electrical and Computer Engineering in the Faculty of Engineering of the University of Porto, Portugal, in 2013 and 2015, respectively.
From 2016 to 2019, he worked as Analog Design Engineer for Synopsys, where he worked on the design and verification of mixed-signal circuits for High-Speed SERDES.
Since 2019, he has been pursuing a Ph.D. degree with the Institute of Neuroinformatics, University of Zürich and ETH Zürich, Switzerland. 
His current research interests include neuromorphic engineering and low noise design for event-based sensors.
\end{IEEEbiography}

\begin{IEEEbiography}
[{\includegraphics[width=1in,height=1.25in,clip,keepaspectratio]
{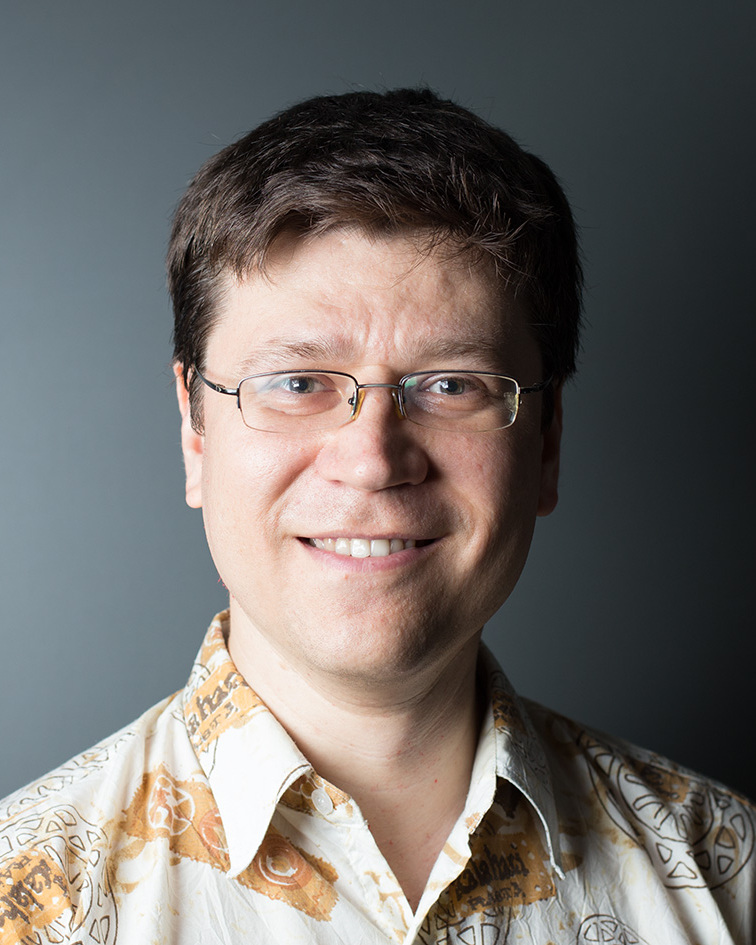}}]
{Ilya Kiselev}
(Member, IEEE) received the Specialist degree in physics from Tambov State University, Russia, in 2000, the M.Sc. degree in applied mathematics and physics from the Moscow Institute of Physics and Technology in 2002, and the Ph.D. degree from ETH Zürich in 2021. He is currently doing his post-doctoral work at the Institute of Neuroinformatics, University of Zürich and ETH Zürich. His research interests include hardware implementations of signal acquisition and processing for traditional and event-based audio processing.
\end{IEEEbiography}

\begin{IEEEbiography}
[{\includegraphics[width=1in,height=1.25in,clip,keepaspectratio]
{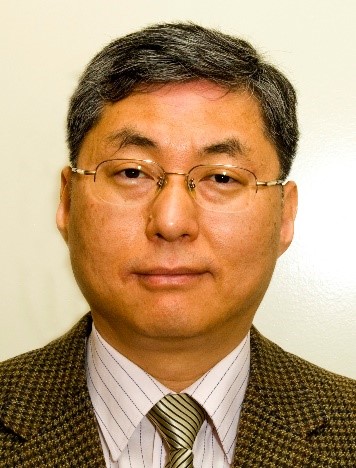}}]
{Hoi-Jun Yoo}
(Fellow, IEEE) graduated from the Department of Electronics, Seoul National University, Seoul, South Korea, in 1983, and the M.S. and Ph.D. degrees in Electrical Engineering from the Korea Advanced Institute of Science and Technology (KAIST), Daejeon, South Korea, in 1985 and 1988, respectively.

He served as a member for the Executive Committee of ISSCC, Symposium on Very Large-Scale Integration (VLSI), and Asian Solid-State Circuits Conference (A-SSCC), the TPC Chair for the A-SSCC 2008 and International Symposium on Wearable Computer (ISWC) 2010, the IEEE Distinguished Lecturer from 2010 to 2011, the Far East Chair for the ISSCC from 2011 to 2012, the Technology Direction Sub-Committee Chair for the ISSCC in 2013, the TPC Vice Chair for the ISSCC in 2014, and the TPC Chair for the ISSCC in 2015. More details are available at
http://ssl.kaist.ac.kr.
\end{IEEEbiography}

\begin{IEEEbiography}
[{\includegraphics[width=1in,height=1.25in,clip,keepaspectratio]
{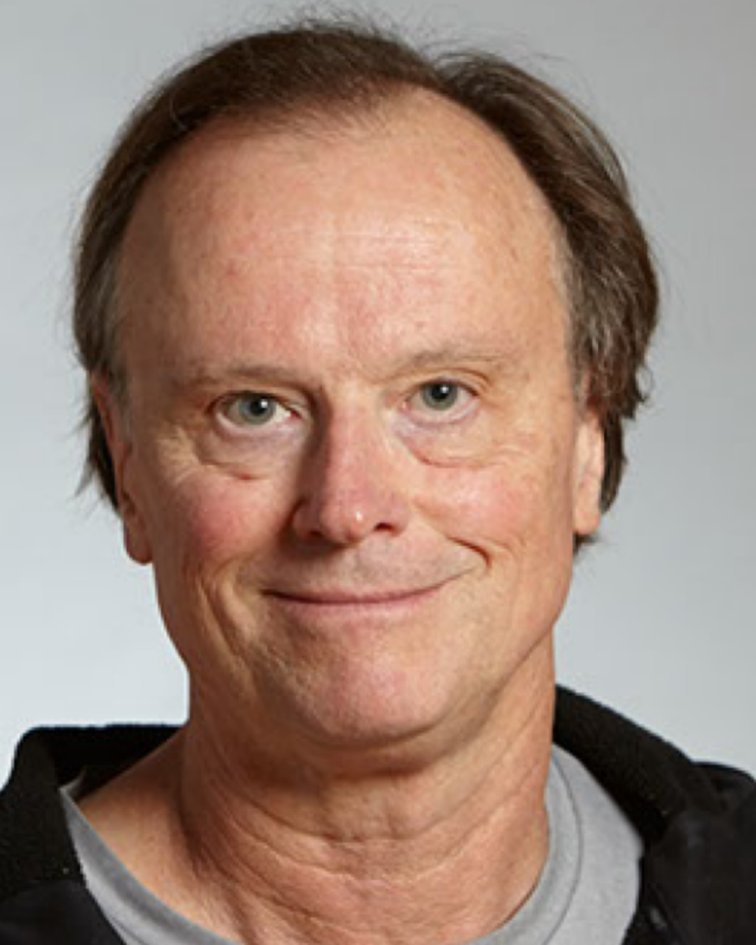}}]
{Tobi Delbruck}
(Fellow, IEEE) received the degree in physics from University of California in 1986 and Ph.D. degree from Caltech in 1993. Currently, he is a Professor of Physics and Electrical Engineering at the Institute of Neuroinformatics, University of Zurich and ETH Zurich, where he has been since 1998. The Sensors group, which he co-directs with Shih-Chii Liu, currently focuses on neuromorphic sensory processing, control, and efficient hardware AI.
\end{IEEEbiography}

\begin{IEEEbiography}
[{\includegraphics[width=1in,height=1.25in,clip,keepaspectratio]
{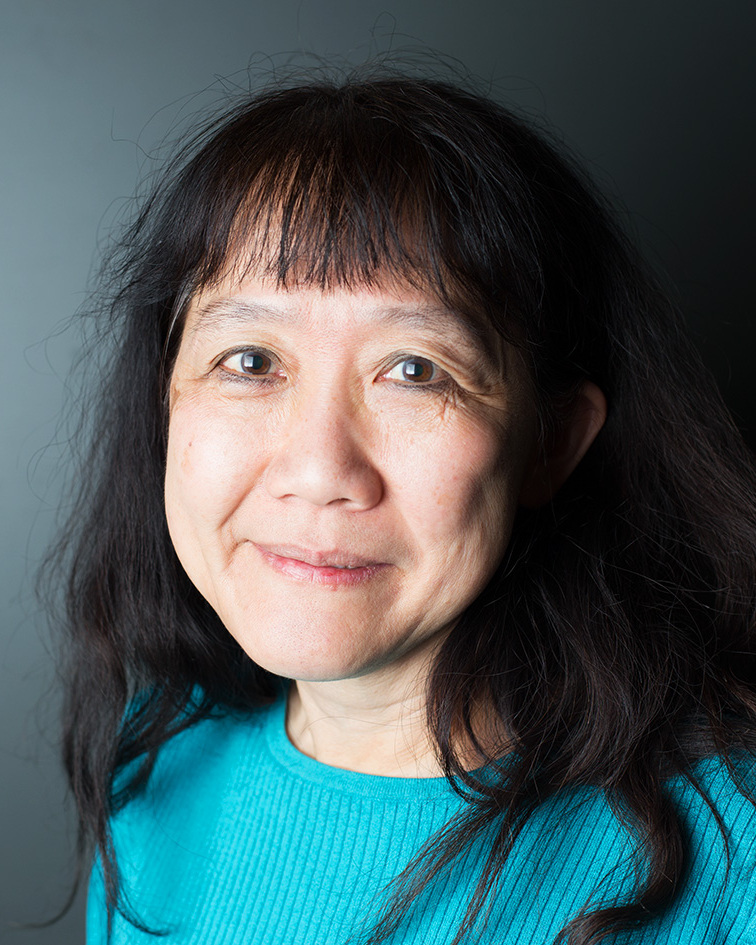}}]
{Shih-Chii Liu}
(Fellow, IEEE) received the bachelor’s degree in electrical engineering from the Massachusetts Institute of Technology, Cambridge, MA, USA, and the Ph.D. degree in the computation and neural systems program from the California Institute of Technology, Pasadena, CA, USA, in 1997. She is currently a Professor at the University of Zurich, Zurich Switzerland. Her group focuses on audio sensors particular the spiking cochlea and bio-inspired deep neural network algorithms and hardware.
\end{IEEEbiography}

\end{document}